\documentstyle[12pt]{article}
\def\a{\alpha}  \def\d{\delta}
\def\b{(\delta -\alpha)}

\def\q2a{q^{2(\a ,\a)}}
\def\q-a{q^{-(\a ,\a)}}
\newcommand{\CW}{Cartan-Weyl }
\newcommand{\Rmt}{ universal $R$-matrix }
\newcommand{\D}{\mbox{$\underline{\Delta} _{+}\,$}}

\newcommand{\Uq}{\mbox{$U_{q}(g)\,$}}

\newcommand{\bn}{\begin{equation}}
\newcommand{\ed}{\end{equation}}

\newtheorem{proposition}{Proposition}[section]

\newtheorem{theorem}{Theorem}[section]
\newcommand{\ea}[1]{\mbox{$e_{{\alpha}_{#1}}$}}

\newcommand{\ema}[1]{\mbox{$e_{{-\alpha}_{#1}}$}}
\newcommand{\epma}[1]{\mbox{$e_{{\pm\alpha}_{#1}}$}}

\newcommand{\hsp}{\mbox{$\hspace{.5in}$}}

\begin{document}

\setcounter{equation}{0}

\title{{\bf Twisting of quantum (super)algebras.
Connection of Drinfeld's and  Cartan-Weyl realizations for quantum
affine algebras}}

\author{{\bf S.M. Khoroshkin}$^{1 \, \, 2}$
and {\bf V.N. Tolstoy}$^{3}$ \\
Max-Planck Institut f\"{u}r Mathematik,\\
Bonn, Germany}
\date{}
\maketitle
\begin{abstract} We show that some factors of the universal R-matrix
generate a family of twistings for the standard Hopf structure of any
quantized contragredient Lie (super)algebra of finite growth.
As an application we prove that any two isomorphic superalgebras with
different Cartan matrices have isomorphic q-deformations (as associative
superalgebras) and their standard comultiplications are connected
by such twisting. We present also an explicit relation between the
generators of the second Drinfeld's realization and Cartan-Weyl
generators of quantized affine nontwisted Kac-Moody algebras.
Further development of the theory of quantum Cartan-Weyl basis,
closely related with this isomorphism, is discussed.
We show that Drinfeld's formulas of a comultiplication for the second
realization are a twisting of the standard comultiplication by  factors
of the universal R-matrix.  Finally, properties of the Drinfeld's
comultiplication are considered.
\end{abstract} \vspace{1cm} \vspace{\fill}

\noindent
MPIM BONN, MPI/94-23
\vspace{12 mm}

\noindent
$^{1}$ Permanent address: ITEP, Bol'shaya Cheremushkinskaya 25,
117259 Moscow, Russia. E-mail: $<$khor@s43.msk.su$>$. \\
$^{2}$ Partially supported by AMS FSU foundation. \\
$^{3}$ Permanent address: Institute of Nuclear Physics, Moscow State
University, 119899 Moscow, Russia. E-mail: $<$tolstoy@compnet.msu.su$>$.

\newpage
\parskip= 4pt plus 1pt

\setcounter{equation}{0}
\section{Introduction}
Numerous applications of quantum algebras are based on the fact that
quantized enveloping algebras have nontrivial algebraic and coalgebraic
structures as Hopf algebras. In addition this gives a possibility to use
for their study not only  automorphisms of the algebras but also
twistings of their coalgebraic structure, which do not change the
structure of multiplication.

\noindent
More generally, first notion of twisting for
quasi-Hopf algebras was introduced and successfully applied in
classification theorems by V. Drinfeld ~\cite{D1}. N. Reshetikhin
\cite{R} observed that one can use, analogously to \cite{D1},
a two-tensor $F \in U_q(g)\otimes U_q(g)$ as a twisting operator and
obtain as a result a new Hopf algebra (without 'quasi' prefix) if $F$
satisfies some natural conditions.
He showed also that multiparameter deformations of the quantum
enveloping algebra $U_{q}(g)$ of a simple Lie algebra $g$ can be
constructed by twisting of $U_{q}(g)$ with the two-tensor $F$
depending on Cartan subalgebra of $U_{q}(g)$.
Such type of twisting was used by Ya. Soibelman [S] for quantization
of the Lie-Poisson structure in compact Lie groups.
Another type of the twisting was considered by B. Enriquez [E].
He showed that the usual (non-deformed) enveloping algebra $U(g)$
of a simple Lie algebra $g$ can be done noncocommutative by a twisting
while the algebraic structure is not changed.
Such a deformation of the coalgebraic sector can be also defined
for some non-semisimple algebras.
Recently it was shown in  [LNRT] that the universal enveloping
algebra of the classical Poincare algebra admits a family of twistings
of its coalgebraic sector without changing of the algebraic sector.

We consider here twistings of quantized contragredient Lie
(super)algebras of finite growth. (These (super)algebras are q-analogs
of all finite-dimensional simple Lie algebras, classical superalgebras
and of all infinite-dimensional affine Kac-Moody (super)algebras). All
these quantum (super)algebras $U_{q}(g)$ are quasitriagular, i.e. they
have the universal R-matrix. Explicit formula for the universal R-matrix
is described by a product of factors over a positive root system of
a Lie (super)algebra. We show that the factors of the universal R-matrix
define a family of twistings for $U_q(g)$ and demonstrate their
connection with other twistings induced Lusztig automorphisms
\cite{KacdeConchini}. This is known in mathematical folklore for
Drinfeld-Jimbo deformations of the simple finite-dimensional Lie
algebras. In other cases we prove as consequences the following
important results:

\noindent
(i) We exhibit a connection between Drinfeld-Jimbo quantizations
of two isomorphic contragredient Lie superalgebras $g$ and $g'$.
More precisely, we show that there exists an isomorphism $\omega:
U_q(g')\mapsto U_q(g)$ of these algebras (the superanalog of
the Lusztig automorphism [L], \cite{KacdeConchini}),
and the standard comultiplications of $U_q(g)$ and  $U_q(g')$ commute
with $\omega$ modulo twisting by corresponding factors of the universal
R-matrix for $U_q(g)$ or $U_{q}(g')$ (see \cite{KT1}).

\noindent
(ii) We present a detailed study of the second Drinfeld's realization
[D2] of the quantum affine algebras by considering its Cartan-Weyl
basis and by studying its twisting. We write down an explicit relation
between generators of the second Drinfeld's realization and the
Cartan-Weyl generators for the quantized affine nontwisted Kac-Moody
algebras (see [KT2] and also \cite{DF} for $\hat{gl}_n$ case).
It turns out that this relation has the most natural form in terms
of the ''circular'' Cartan-Weyl generators, introduced in \cite{KT3}.
These generators realize the idea of natural extension of the normal
ordering from the positive root system to the set of all root.
It should be noted that analogous construction appears in
a Hall algebra approach to a construction of the Cartan-Weyl
generators in terms of a quiver \cite{Ringel} if we wish to extend
this construction to a derived category of representations of
the quiver.

\noindent
(iii) We show that Drinfeld's formulas describing the comultiplication
in the second realization are obtained by a twisting of the standard
comultiplication by a factor of the universal R-matrix.  This twisting
is correctly defined for an appropriate completion of
$U_q(\hat{g})\otimes U_q(\hat{g})$, and corresponds to a "virtual"
longest element $\omega_0$ of the affine q-Weyl group.
The origin of such construction one can see on the quasiclassical level
where $\omega_0$ does not act on elements of a Lie algebra but
interchange Manin triples which are responsible for two different
quantizations of a current algebra.

\noindent
(iv) Finally, we discuss also some properties of the natural
comultiplication in the second realization of the quantum affine
algebras \cite{D3}. Unfortunately, this comultiplication is still
not commonly used. We demonstrate the meaning of the
quasicocommutativity condition for this comultiplication,
then we present the universal R-matrix and show that for concrete
representations this universal R-matrix produces solutions of the
Yang-Baxter equation with entries being generalized functions of
a spectral parameter.

The paper is organized as follows. In Section 2 we remind the definition
of any quantized finite-dimensional contragredient Lie (super)algebra
$g$ (or a quantum (super)algebra $U_{q}(g)$) in terms of the Chevalley
generators and a q-(super)commutator and also in terms of the quantum
adjoint action.

\noindent
In Section 3 we present the method of construction of the quantum
\CW basis, its ''circular'' generalization and define  some extensions
of  $U_{q}(g)$ and $U_{q}(g) \otimes U_{q}(g)$ which we need for the
definitions of twisting and of the universal R-matrix. The explicit
formula for the universal R-matrix is presented in the Section 4.

\noindent
In Section 5 we discuss at first some general properties of  twistings
for an arbitrary Hopf (super)algebra, then we consider the twistings
by factors of the universal R-matrix.

\noindent
The Sections 6-8 are devoted to the second Drinfeld's realization of
the quantized affine algebras.

\noindent
In Appendices A, B one can find some details of the considerations for
the case of $U_{q}(\hat{sl}_2)$.

\setcounter{equation}{0}
\section{Quantized Lie (super)algebras of finite growth}

Let $g(A,\Upsilon)$ be any contragredient Lie (super)algebra of finite
growth with a symmetrizable Cartan  matrix $A$ ($A^{sym}=(a_{ij}^{sym})$
is a corresponding symmetrical matrix) and let $\Pi:=\{\alpha_{1},
\ldots , \alpha_{r}\}$ be a system of simple roots for $g(A,\Upsilon)$
\footnote{These (super)algebras are all  finite-dimensional simple Lie
(super)algebras and all infinite-dimensional affine Kac-Moody
(super)algebras [K1].}. The quantized (super)algebra $g:=g(A,\Upsilon)$
is an unital associative (super)algebra \Uq
with the Chevalley generators $e_{\pm \alpha_{i}},\;
k_{\alpha_{i}}^{\pm 1}=
q^{\pm h_{\alpha_{i}}},\; (i \in {\rm I}:=\{1,2,\ldots,r\})$, and
the defining relations [T1,KT1,KT3,KT4] \
\bn
[k_{\alpha_{i}}^{\pm 1}, k_{\alpha_{j}}^{\pm 1}]=0 \ ,\hsp
k_{\alpha_{i}}e_{\pm \alpha_{j}}= q^{\pm (\alpha_{i},\alpha_{j})}
e_{\pm\alpha_{j}} k_{\alpha_{i}} \\ ,
\label{Q1}
\ed
\bn
[e_{\alpha_{i}}, e_{-\alpha_{j}}] = \delta_{ij}\frac{k_{\alpha_{i}} -
k_{\alpha_{i}}^{-1}}{q-q^{-1}} \ ,
\label{Q2}
\ed\
\bn
(\tilde{\rm ad}_{q'}\epma{i})^{n_{ij}+1}\epma{j}=0 \hsp
{\rm for}\;\; i\neq j,\;\; q'=q,q^{-1} \ ,
\label{Q3}
\ed\
\[
\deg(k_{\alpha_{i}})=\deg(\epma{j})= \bar{0}  \hsp
{\rm for}\;\; i \in {\rm I} \ , \;\; j \not \in \Upsilon \ ,
\]
\bn
\deg(\epma{i})= \bar{1} \hsp {\rm for}\;\; i \in  \Upsilon \subset
{\rm I} \ , \label{Q4}
\ed
where
\bn
n_{ij}=\left\{\begin{array}{lll}
\;0 \;\;\;\; {\rm if}\; a_{ii}^{sym}=a_{ij}^{sym}=0 \ , \\\
1 \;\;\;\; {\rm if}\; a_{ii}^{sym}=0,\;a_{ij}^{sym}\neq 0 \ , \\\
-2(a_{ij}^{sym}/a_{ii}^{sym})\;\;\;\; {\rm if}\; a_{ii}^{sym}\neq 0 \ .
\end{array}\right .
\label{Q5}
\ed
Moreover, there are the following additional triple relations [KT1,KT3]\
\bn
{[{[\epma{j},\epma{i}]}_{q'},{[\epma{j},\epma{l}]}_{q'}]}_{q'}=0 \ ,
\hsp {\rm for}\,\,\,\, q'=q,\,\,q^{-1} \ ,
\label{Q6}
\ed
if the three simple roots $\alpha_{i},\alpha_{j},\alpha_{l}\in \Pi$
satisfy the condition
\bn
(\alpha_{j},\alpha_{j})=(\alpha_{i},\alpha_{l})=(\alpha_{j},
\alpha_{i}+\alpha_{l})=0 \ .
\label{Q7}
\ed
Here in (\ref{Q1})-(\ref{Q3}), (\ref{Q6}) the bracket ${[\cdot,\cdot]}$
is an usual supercommutator, $\tilde{\rm ad}_{q'}$ and
${[\cdot ,\cdot ]}_{q}$ denote a deformed supercommutator
(q-supercommutator) in \Uq :\
\bn
(\tilde{\rm ad}_{q'}e_{\alpha})e_{\beta} \equiv
{[e_{\alpha},e_{\beta}]}_{q'}=
e_{\alpha}e_{\beta}-(-1)^{\theta (e_{\alpha})\theta (e_{\beta})}
q'^{(\alpha ,\beta )}e_{\beta}e_{\alpha} \ ,
\label{Q8}
\ed
where $(\alpha ,\beta )$ is a scalar  product of the roots $\alpha$
and $\beta$: $(\alpha_{i},\alpha_{j})=a_{ij}^{sym}$.
In the formula (\ref{Q8}) and below we use the short notation
\bn
\theta (\gamma):=\theta (e_{\gamma})\equiv \deg (e_{\gamma}) \ .
\label{Q9}
\ed
{\it Remarks.}  (i) The triple relations (\ref{Q6}) may appear
only in the supercase for the following situation in the Dynkin diagram:
\bn
\setlength{\unitlength}{0.7mm}
     \begin{picture}(180,15)(-80,0)
     \put(-30,8){\makebox(0,0){$\alpha_{i}$}}
     \put(0,8){\makebox(0,0){$\alpha_{j}$}}
     \put(30,8){\makebox(0,0){$\alpha_{l}$}}
     \put(-30,0){\circle*{0.8}}
     \put(0,0){\makebox(0,0){$\otimes$}}
    \put(30,0){\circle*{0.8}}
     \put(-25,0){\makebox{\line (1,0){20}}}
     \put(5,0){\makebox{\line (1,0){20}}}
     \end{picture}
\label{Q10}
\ed
where $\alpha_{j}$ is a grey root and the roots $\alpha_{i}$ and
$\alpha_{l}$ are not connected and they can be of any color: white,
grey or dark.

\noindent
(ii) The outer q-supercommutator in (\ref{Q6}) is actually an usual
one since $(\alpha_{i}+\alpha_{j}, \alpha_{j}+\alpha_{l})=0$.

\noindent
(iii) The triple relations have evident classical counterparts.

The quantum (super)algebra \Uq is a Hopf (super)algebra with respect
to a comultiplication $\Delta _{q'}$, an antipode $S_{q'}$ and a counit
$\varepsilon$ defined as
\bn
\Delta _{q'}(k_{\alpha_{i}})=k_{\alpha_{i}}\otimes k_{\alpha_{i}} \ ,
\label{Q11}
\ed
\bn
\Delta _{q'}(e_{\alpha_{i}}) = e_{\alpha_{i}}\otimes 1+
k'_{\alpha_{i}} \otimes e_{\alpha_{i}} \ ,
\label{Q12}
\ed
\bn
\Delta _{q'}(e_{-\alpha_{i}}) = e_{-\alpha_{i}}\otimes
k_{\alpha_{i}}'^{-1} + 1 \otimes e_{-\alpha_{i}} \ ,
\label{Q13}
\ed\
\[
S_{q'}(k'_{\alpha_{i}})=k_{\alpha_{i}}'^{-1} \ ,
\]
\bn
S_{q'}(\ea{i})=-k_{\alpha_{i}}'^{-1}\ea{i} \ ,
\hsp S_{q'}(\ema{i})=-\ema{i}k'_{\alpha_{i}} \ ,
\label{14}
\ed\
\bn
\varepsilon (k_{\alpha_{i}})=\varepsilon (\ea{i})=
\varepsilon (\ema{i})= 0 \ ,\hsp \varepsilon (1)=1 \ ,
\label{Q15}
\ed
where $k_{\alpha}'=q'^{h_{\alpha}}$ and $q'$ may be chosen
as $q'=q$ or $q'=q^{-1}$.

We may rewrite the defining relation by means of an adjoint action
of \Uq on itself. For this aim we introduce new  Chevalley
generators $\hat{e}_{\pm \alpha_i}$ by the following formulas
\bn
\hat{e}_{\alpha_i}=\ea{i}, \hsp
\hat{e}_{-\alpha_i}=q'^{-1}\ema{i} k'_{\alpha_{i}} \ .
\label{Q16}
\ed
In this basis the relations (\ref{Q2}), (\ref{Q3}), (\ref{Q6})
take the following form [KT3]
\bn
({\rm ad}_{q'}\hat{e}_{\alpha_i}) \hat{e}_{-\alpha_j}=
{[\hat{e}_{\alpha_i},\hat{e}_{-\alpha_j}]}_{q'}=
\delta_{ij}\frac{1-k_{\alpha_{i}}'^{2}}{1-q'^{2}} \ ,
\label{Q17}
\ed
\bn
{({\rm ad}_{q'}\hat{e}_{\pm \alpha_i})}^{n_{ij}+1}\hat{e}_{\pm \alpha_j}=
0 \ , \hsp (i\neq j) \ ,
\label{Q18}
\ed
\bn
[({ \rm ad}_{q'}\hat{e}_{\pm\alpha_j})\hat{e}_{\pm\alpha_i},
({\rm ad}_{q'}\hat{e}_{\pm\alpha_j})\hat{e}_{\pm\alpha_{l}}]=0 \ .
\label{Q19}
\ed
The last relation holds for the condition (\ref{Q7}).
Here ${\rm ad}_{q'}$ is the quantum adjoint action (see details in [KT3])
defined by \
\bn
({\rm ad}_{q'}\,a)x:=({\rm id}\otimes S_{q'})\Delta_{q'}(a))\circ x
\label{Q20}
\ed
for all homogeneous elements $a,x\in \Uq$, where the operation $\circ$
is defined by the rule
\bn
(a\otimes b)\circ x=(-1)^{\theta(b)\theta(x)}axb \ .
\label{Q21}
\ed

Below we denote by a symbol $(^{*})$ an anti-involution in $U_{q}(g)$,
defined as $(k_{\alpha_{i}})^{*}=k_{\alpha_{i}}^{-1},\;$
$(e_{\pm\alpha_{i}})^{*}=e_{\mp\alpha_{i}},\;$ $(q)^{*}=q^{-1}$.
We also use the standard notations $U_{q}(\kappa)$ and $U_{q}(b_{\pm})$
for the Cartan and Borel subalgebras, generated by
$k_{\alpha_{i}}^{\pm 1}$
and $e_{\pm\alpha_{i}}$, $k_{\alpha_{i}}$, $k_{\alpha_{i}}^{-1}$
correspondingly. We write also
\bn
{\rm exp}_{q}(x) := 1 + x + \frac{x^{2}}{(2)_{q}!} + \ldots +
\frac{x^{n}}{(n)_{q}!} + \ldots =
\sum_{n\geq 0} \frac{x^{n}}{(n)_{q}!} \ ,
\label{Q22}
\ed
\bn
(a)_{q}:=\frac{q^{a}-1}{q-1} \ ,\,\,\,\,\,\,\,\,\,\,\,\,
[a]_{q}:=\frac{q^{a}-q^{-a}}{q-q^{-1}} \ ,\,\,\,\,\,\,\,\,\,\,\,\,
q_{\alpha}:=(-1)^{\theta (\alpha)}q^{(\alpha,\alpha)} \ .
\label{Q23}
\ed

Now we proceed to a description of the Cartan-Weyl basis
for the quantum (super)algebras $U_{q}(g)$.

\setcounter{equation}{0}
\section{Cartan-Weyl basis for $U_{q}(g)$ and its generalizations}

Let $\Delta _{+}$ be the system of all positive roots for
$g(A,\Upsilon)$
with respect to $\Pi$. We denote by $\underline{\Delta}_{+}$
{\it the reduced root system} which is obtained from $\Delta _{+}$
by removing such roots $\alpha$ for which $\alpha /2$ are {\it odd}
roots.

Our procedure of a construction of the quantum Cartan-Weyl basis for
$U_{q}(g)$ is in agreement with a choice of a normal ordering
in $\underline{\Delta}_{+}$. We remind the definition of the normal
ordering in $\underline{\Delta}_{+}$ [AST,T2,T3].

\noindent
{\it We say that the system \D is in the normal ordering if each
composite root $\gamma=\alpha + \beta \in \underline{\Delta}_{+}$,
where $\alpha \neq \lambda\beta$,
$\alpha$, $\beta \in \underline{\Delta}_{+}$, is written between its
components $\alpha$ and $\beta$.}

\noindent
It should be noted that for any finite-dimensional simple Lie algebra
there is one-to-one correspondence between the normal orderings and
reduced decompositions of the longest element of the Weyl group [Z]
(see Section 5). We have no such correspondence for the Lie
superalgebras and the affine Lie algebras because the superalgebras
have no  any "good" Weyl group and the affine algebras have not any
longest element of the Weyl group.

\noindent
{\it We shall say  that $\alpha < \beta$ if $\alpha$ is located on the
left of $\beta$ in the normal ordering system $\underline{\Delta}_{+}$}.

\noindent
The quantum Cartan-Weyl basis is being constructed by using
the following inductive algorithm [T1,KT1,KT3,KT4,TK].

\noindent
{\bf Algorithm 3.1} {\it We fix some normal ordering in \D and put by
induction
\bn
e_{\gamma} := [e_{\alpha},e_{\beta}]_{q} \ ,\,\,\,\,\,\,\,
e_{-\gamma} := [e_{-\beta},e_{-\alpha}]_{{q}^{-1}}
\label{CW1}
\ed
\noindent if $\gamma=\alpha+\beta$, $\alpha<\gamma<\beta$, and
$[\alpha;\beta]$ is a minimal segment including $\gamma$,
i.e. the segment has not other roots  $\alpha'$ and $\beta'$
such that $\alpha'+\beta'=\gamma$. Moreover we put\
\bn
k_{\gamma} :=\prod_{i=1}^{r} k_{\alpha_{i}}^{l_{i}}  ,
\label{CW2}
\ed
if $\gamma = \sum_{i=1}^{r}l_{i}\alpha_{i}$}, ($ \alpha_i \in \Pi$).

\noindent
By this procedure one can construct the total quantum Cartan-Weyl basis
for all finite-dimensional contragredient simple (super)algebras.
In the case of the infinite-dimensional affine Lie (super)algebras
we use an additional constraint. Namely, we construct at first
all root vectors by our procedure and then we redefine the root
vectors of the imaginary roots so that new imaginary root vectors
commute if they are not conjugated. That is, e.g., let
$e_{\pm n\delta}'^{(i)}$ be root vectors of the imaginary roots
$\pm n\delta$ \footnote{In the case of a quantum affine algebra
$U_{q}(g)$ the root vectors of the imaginary roots $\gamma =\pm n\delta$
have to be labeled by an additional index $i$:
$e_{\pm n\delta}'^{(i)}$, $i = 1,2,\ldots, {\it mult}$,
where $mult$ is a multiplicity of the imaginary root $\pm n\delta$.},
constructed by the procedure.
It turns out that
\bn
[e_{n\delta}'^{(i)}, e_{-m\delta}'^{(j)}] \neq \delta_{m,-n}
a_{ij}(n) \frac{k^{n}_{\delta}-k^{-n}_{\delta}}{q - q^{-1}} \ .
\label{CW3}
\ed
We introduce the new vectors $e_{\pm n\delta}^{(i)}$:
\bn
e_{\pm n\delta}^{(i)} =
p(e_{\pm \delta}'^{(i)},e_{\pm 2\delta}'^{(i)} \ ,
\ldots, e_{\pm n\delta}'^{(i)})
\label{CW4}
\ed
which will satisfy the relation
\bn
[{e}_{n\delta}^{(i)}, {e}_{-m\delta}^{(j)}] = \delta_{m,-n}
a_{ij}(n)\frac{k^{n}_{\delta}-k^{-n}_{\delta}}{q - q^{-1}} \ .
\label{CW5}
\ed
This relation agrees with its classical counterpart.

The quantum Cartan-Weyl generators constructed by the procedure are
characterized by the following basic properties [T1,KT1,KT3].
\begin{theorem}
The root vectors $e_{\pm\gamma} \in U_{q}(g)$ and the Cartan elements
$k_{\gamma} \in U_{q}(g)$ for all $\gamma\,\in \D$
satisfy the following relations:
\bn
(e_{\pm\gamma})^{*}=e_{\mp\gamma} \ , \hsp
k_{\alpha}^{\pm 1}e_{\gamma} = q^{\pm(\alpha,\gamma)}e_{\gamma}
k_{\alpha}^{\pm 1} \ ,
\label{CW6}
\ed
\bn
[e_{\gamma},e_{-\gamma}] = a(\gamma)\frac{k_{\gamma}-k_{\gamma}^{-1}}
{q-q^{-1}} \ ,
\label{CW7}
\ed
\bn
{[e_{\alpha},e_{\beta}]}_{q} = \sum_{\alpha <\gamma _{1}< \ldots
<\gamma_{m} < \beta} C_{n_{j},\gamma _{j}} e_{\gamma _{1}}^{n_{1}}
e_{\gamma _{2}}^{n_{2}}\cdots e_{\gamma _{m}}^{n_{m}} \ ,
\label{CW8}
\ed
for $\alpha, \beta \in \D$, where $\sum_{j} n_{j}\gamma _{j} =
\alpha + \beta$, and the coefficients
$C_{\ldots}$ are rational functions of $q$ and ones do not depend on
the Cartan elements $k_{\alpha_{i}},\: i=1,2,\ldots r$. Moreover
\bn
{[e_{\beta },e_{-\alpha}]}=
\sum C'_{n_{j},\gamma_{j};n'_{j},\gamma'_{j}}e_{-\gamma_{1}}^{n_{1}}
e_{-\gamma_{2}}^{n_{2}}\cdots e_{-\gamma _{m}}^{n_{m}}
e_{\gamma'_{1}}^{n'_{1}}e_{\gamma'_{2}}^{n'_{2}}\cdots
e_{\gamma'_{l}}^{n'_{l}}
\label{CW9}
\ed
\noindent
where the sum is taken on $\gamma_{1},\ldots,\gamma_{m},\gamma'_{1},
\ldots ,\gamma'_{l}$  and $n_{1},\ldots,n_{m}, n'_{1},\ldots,n'_{l}$
such that
\[
\gamma_{1}< \ldots <\gamma_{m} <\alpha <\beta <\gamma'_{1} <\ldots
<\gamma'_{l} \ ,
\]
\[
\sum_{j}(n'_{j}\gamma'_{j} - n_{j}\gamma_{j}) = \beta -\alpha \ ,
\]
\noindent
and the coefficients $C'_{\ldots} $ are rational functions of $q$ and
$k^{-1}_{\alpha}$ or $k^{-1}_{\beta}$. The monomials
$e_{\gamma_{1}}^{n_{1}}
e_{\gamma_{2}}^{n_{2}} \cdots e_{\gamma_{m}}^{n_{m}}$ and
$e_{-\gamma_{1}}^{n_{1}}e_{-\gamma_{2}}^{n_{2}}\cdots
e_{-\gamma_{m}}^{n_{m}}$, ($\gamma_{1} <\gamma_{2}<\cdots <\gamma_{m}$),
generate (as a linear space over $U_{q}(\kappa)$) the
subalgebras $U_{q}(b_{+})$ and $U_{q}(b_{-})$ correspondingly.
The monomials
\bn
e_{-\gamma_{1}}^{n_{1}}e_{-\gamma_{2}}^{n_{2}}\cdots
e_{-\gamma_{m}}^{n_{m}}e_{\gamma'_{1}}^{n'_{1}}
e_{\gamma'_{2}}^{n'_{2}}\cdots e_{\gamma'_{l}}^{n'_{l}} \ ,
\label{CW10}
\ed
where $\gamma_{1}<\gamma_{2} <\cdots <\gamma_{m}$ and
$\gamma'_{1} <\gamma'_{2} <\cdots <\gamma'_{l}$), generate $U_{q}(g)$
over $U_{q}(\kappa)$.
\end{theorem}

\noindent
If there are imaginary root vectors in the relations
(\ref{CW6})-(\ref{CW9}) then we should use an additional index for
such vectors. For example, the relation (\ref{CW7}) for
$\gamma = \pm n \delta$ has the form (\ref{CW5}).

We can transform the root vectors $e_{\pm\gamma}$ in new ones such that
the coefficients $C_{\ldots}$ and $C'_{\ldots}$  in (\ref{CW8}) and
(\ref{CW9}) will not depend on the Cartan elements $k_{\gamma}$.
For this goal we extend the notation of the normal ordering for \D
to a "circular" (or "clockwise") normal ordering for the reduced system
of all roots, $\underline{\Delta}:=\D\cup(-\D)$.

\noindent
Let $\gamma_{1}, \gamma_{2}, \ldots, \gamma_{N}$ be a normal ordering
in \D then a circular normal ordering in $\underline{\Delta}$ means that
the roots of $\underline{\Delta}$ are located on a circular by
the following way (see [KT3])
\bn
\gamma_{1}, \gamma_{2}, \ldots, \gamma_{N}, -\gamma_{1}, -\gamma_{2},
\ldots, -\gamma_{N}, \gamma_{1}
\label{CW11}
\ed
or

\setlength{\unitlength}{0.6mm}
     \begin{picture}(180,120)(-80,0)
     \put(0,50){\makebox(0,0){$\gamma_1$}}
     \put(4,69){\makebox(0,0){$\gamma_2$}}
     \put(15,85){\circle*{1.5}}
     \put(31,96){\circle*{1.5}}
     \put(50,100){\circle*{1.5}}
     \put(69,96){\circle*{1.5}}
     \put(85,85){\circle*{1.5}}
     \put(96,69){\makebox(0,0){$\gamma_N$}}
     \put(100,50){\makebox(0,0){$-\gamma_1$}}
     \put(96,31){\makebox(0,0){$-\gamma_2$}}
     \put(85,15){\circle*{1.5}}
     \put(69,4){\circle*{1.5}}
     \put(50,0){\circle*{1.5}}
     \put(31,4){\circle*{1.5}}
     \put(15,15){\circle*{1.5}}
     \put(4,31) {\makebox(0,0){-$\gamma_N$}}
     \put(64,88){\vector (3,-2){10}}
     \put(36,12){\vector (-3,2){10}}
     \end{picture}
\vspace{0.4cm} \[\mbox{Figure 1}\] \vspace{0.2cm}

\noindent
{\it We say that a segment $[\gamma', \gamma'']$ of the circle,
where $\gamma'$ and $\gamma''$ are any roots of $\underline{\Delta}$
and the root $\gamma''$ follows after $\gamma'$ in accordance with
the given direction on the circle, is minimal if it does not contain
the opposite roots $-\gamma'$ and $-\gamma''$.

 \noindent
 We say also that $\gamma' < \gamma''$ if the segment
 $[\gamma',\gamma'']$ is minimal.}

\noindent
We introduce two type of circular root vectors $\hat{e}_{\gamma}$
and $\check{e}_{\gamma}$ by the following formulas:
\bn
\hat{e}_{\gamma}:= e_{\gamma} \ , \hsp
\hat{e}_{-\gamma}:=-k_{\gamma}^{-1} e_{-\gamma} \ ,
\hsp \forall\,\gamma \in \D \ ,
\label{CW12}
\ed
and
\bn
\check{e}_{-\gamma}:= e_{-\gamma} \ , \hsp
\check{e}_{\gamma}:= - e_{\gamma} k_{\gamma} \ ,
\hsp \forall\,\gamma \in \D \ .
\label{CW13}
\ed
In terms of these circular generators the relations (\ref{CW8}) and
(\ref{CW9}) are rewritten in the united form\
\bn
{[\hat{e}_{\alpha},\hat{e}_{\beta}]}_{q} = \sum_{\alpha <\gamma _{1}
< \ldots <\gamma_{m} < \beta} C_{n_{j},\gamma _{j}}
\hat{e}_{\gamma _{1}}^{n_{1}} \hat{e}_{\gamma _{2}}^{n_{2}}
\cdots \hat{e}_{\gamma _{m}}^{n_{m}} \ ,
\label{CW14}
\ed
if $\alpha \in \D$ and  $\alpha < \beta$ in a sense of the circular
normal ordering. We have also\
\bn
{[\check{e}_{\alpha}, \check{e}_{\beta}]}_{q} =
\sum_{\alpha <\gamma _{1}< \ldots <\gamma_{n} < \beta}
C'_{m_{j},\gamma _{j}} \check{e}_{\gamma _{1}}^{m_{1}}
\check{e}_{\gamma _{2}}^{m_{2}}\cdots
\check{e}_{\gamma _{n}}^{m_{n}} \ ,
\label{CW15}
\ed
if $-\alpha \in \D$ and  $\alpha < \beta$ in a sense of the circular
normal ordering. The coefficients $C_{\ldots}$ and $C'_{\ldots}$
in (\ref{CW14}) and (\ref{CW15}) are rational functions of $q$ and
do not depend on the elements $k_{\gamma}$.

\noindent
It should be noted that we can construct the circular root vectors
$\hat{e}_{\pm\gamma}$ (up to scalar factors) applying the q-commutator
algorithm to the Chevalley elements $\hat{e}_{\pm \alpha_{i}}$.

\noindent
{\it Remark.} A realization of $U_q(n_+)$ in terms of Hall algebra
for a category ${\cal C}(g)$ of representations of the Dynkin graph
provides a construction of the Cartan-Weyl basis for $U_q(n_+)$
\cite{Ringel} quite analogous to one presented in this section.
In this approach the generators $\hat{e}_\gamma$ and $\check{e}_\gamma$
naturally appear as the Cartan-Weyl generators for different
subcategories of the derived category  of ${\cal C}(g)$.
We are thankful to Prof. C.M. Ringel for discussions of this subject.

Now we want to consider some extensions of $U_{q}(g)$,
$U_{q}(b_{+})\otimes U_{q}(b_{-})$, $U_{q}(g)\otimes U_{q}(g)$ since,
for example, the universal R-matrix is element of two the last
extensions.

\noindent
Let $\mbox{Fract}\, (U_{q}(\kappa))$ be a field of fractions
over $U_{q}(\kappa)$, i.e. $\mbox{Fract}\, (U_{q}(\kappa))$ is
an associative algebra of rational functions of the elements
$k_{\alpha_{i}}^{\pm 1}$, $(i=1,2,\ldots,r)$. Let us
construct a formal Taylor series of the following monomials
\bn
e_{-\beta}^{n_{\beta}} \cdots e_{-\gamma}^{n_{\gamma}}
e_{-\alpha}^{n_{\alpha}} \:  e_{\alpha}^{m_{\alpha}}
e_{\gamma}^{m_{\gamma}} \cdots e_{\beta}^{m_{\beta}}
\label{CW16}
\ed
with coefficients from $\mbox{Fract}\, (U_{q}(\kappa))$, where
$\alpha<\gamma<\cdots<\beta$ in a sense of the fixed normal ordering
in $\underline{\Delta}_{+}$ and the nonnegative integers $n_{\beta},
\ldots, n_{\alpha}, m_{\alpha},\ldots, m_{\beta}$ are subjected to
the constraints
\bn
\mid\sum_{\alpha \in \underline{\Delta}_{+}}(n_{\alpha}-m_{\alpha})
c_{i}^{(\alpha)} \mid\leq \mbox{const},\hsp i=1,2,\cdots,r \ .
\label{CW17}
\ed
Here $c_{i}^{(\alpha)}$ are coefficients in a decomposition of the root
$\alpha$ with respect to the system of the simple roots, $\Pi$.
Let $T_{q}(g)$ be a linear space of all such formal series, then this
space  is an associative algebra with respect to a multiplication of
these formal series and it is called the Taylor extension of $U_{q}(g)$
(see {KT4]).

\noindent
Let $\mbox{Fract}\, (U_{q}(\kappa \otimes \kappa))$ be a field of
fractions generated by the following elements $1\otimes k_{\alpha_{i}}$,
$k_{\alpha_{i}}\otimes 1$ and
$q^{h_{\alpha_{i}} \otimes h_{\alpha_{j}}}$,
($i,j=1,2,\ldots,r$). Let us consider a formal Taylor series of
the following monomials
\bn
e_{-\beta}^{n_{\beta}} \cdots e_{-\gamma}^{n_{\gamma}}
e_{-\alpha}^{n_{\alpha}} \otimes  e_{\alpha}^{m_{\alpha}}
e_{\gamma}^{m_{\gamma}} \cdots e_{\beta}^{m_{\beta}}
\label{CW18}
\ed
with coefficients from $\mbox{Fract}\, (U_{q}(\kappa\otimes \kappa))$,
where $\alpha<\gamma<\cdots<\beta$ in a sense of the fixed normal
ordering in \D and the nonnegative integers
$n_{\beta},\ldots, n_{\alpha},m_{\alpha},
\ldots,m_{\beta}$ are subjected to the constraint (\ref{CW17}).
Let $T_{q}(b_{+}\otimes b_{-})$ be a linear space  of
all such formal series. Then this space  is an associative algebra
with respect to a multiplication
of these formal series  and it is called the Taylor extension
of $U_{q}(b_{+})\otimes U_{q}(b_{-})$ (see [KT4]).

\noindent
At last we take a formal Taylor series of the following monomials
\bn
e_{-\beta}^{n_{\beta}} \cdots e_{-\gamma}^{n_{\gamma}}
e_{-\alpha}^{n_{\alpha}} e_{\alpha}^{m_{\alpha}}
e_{\gamma}^{m_{\gamma}} \cdots e_{\beta}^{m_{\beta}} \otimes
e_{-\beta}^{n'_{\beta}} \cdots e_{-\gamma}^{n'_{\gamma}}
e_{-\alpha}^{n'_{\alpha}} e_{\alpha}^{m'_{\alpha}}
e_{\gamma}^{m'_{\gamma}} \cdots e_{\beta}^{m'_{\beta}}
\label{CW19}
\ed
with coefficients from $\mbox{Fract}\, (U_{q}(\kappa\otimes \kappa))$,
where $\alpha<\gamma<\cdots<\beta$ in a sense of the fixed normal
ordering in $\underline{\Delta}_{+}$ and the nonnegative integers
$n_{\beta},\ldots, n_{\alpha},
m_{\alpha},\ldots, m_{\beta}$ and $n'_{\beta},\ldots, n'_{\alpha}$,
$m'_{\alpha},\ldots, m'_{\beta}$ are subjected to the constraints
\bn
\mid\sum_{\alpha \in \Delta_{+}}(n_{\alpha}+n'_{\alpha}-
m_{\alpha}-m'_{\alpha})c_{i}^{(\alpha)}
\mid\leq \mbox{const},\hsp i=1,2,\cdots, r.
\label{CW20}
\ed
Let $T_{q}(g\otimes g)$ be a linear space  of all such formal series,
then this space is an associative algebra with respect to a
multiplication of these formal series and it is called the Taylor
extension of $U_{q}(g)\otimes U_{q}(g)$ (see [KT4]).

\noindent
The following embedding holds [KT4]
\[
T_{q}(g\otimes g)\supset T_{q}(b_{+}\otimes b_{-}),
\]
\bn
T_{q}(g\otimes g) \supset T_{q}(g)\otimes T_{q}(g) \supset
\Delta_{q'}(T_{q}(g)).
\label{CW21}
\ed

Now we introduce a natural topology in the space
$T_{q}(g\otimes g)$, where basic open neighborhoods of zero,
$\Omega_l$, are defined as linear spans of such series generated
the monomials (\ref{CW19}) with the additional constraint
\bn
\sum_{i=1}^{r}\sum_{\alpha \in \Delta_{+}}
(n_{\alpha}+m_{\alpha} +n'_{\alpha}+ m'_{\alpha})c_{i}^{(\alpha)}
\geq l \ .
\label{CW22}
\ed
Such the topology will be called the formal series (FS) topology.

\noindent
For the goals of  Section 8 we introduce also two other topologies
in $U_q(g)\otimes U_q(g)$. Namely, let $\Omega_l^+$ and $\Omega_l^-$ be
linear spans of monomials from $U_q(g)\otimes U_q(g)$ (\ref{CW19})
with the additional constraints
 \bn
\sum_{i=1}^{r}\sum_{\alpha \in \underline{\Delta}_{+}}
(n_{\alpha}+m_{\alpha} -n'_{\alpha}- m'_{\alpha})c_{i}^{(\alpha)}
\geq l
\label{CW23}
\ed
and
\bn
\sum_{i=1}^{r}\sum_{\alpha \in \underline{\Delta}_{+}}
(-n_{\alpha}-m_{\alpha} +n'_{\alpha}+ m'_{\alpha})c_{i}^{(\alpha)}
\geq l
\label{CW24}
\ed
correspondingly. Then  $\Omega_l^+$ and $\Omega_l^-$ generated two
different topologies in  $U_q(g)\otimes U_q(g)$. We denote their formal
completions (or, equivalently, their closures in $T_{q}(g\otimes g)$)
by $T_{q}^+(g\otimes g)$ and $T_{q}^-(g\otimes g)$.

\setcounter{equation}{0}
\section{Universal R-matrix}

Any quantum (super)algebra $U_{q}(g)$ is a non-cocommutative Hopf
(super)algebra which has an intertwining operator called the universal
R-matrix.

By definition [D4], the universal R-matrix for the Hopf (super)algebra
$U_{q}(g)$ is an invertible element $R$ of the Taylor extension
$T_{q}(b_{+}\otimes b_{-})$, satisfying the equations
\bn
\tilde{\Delta}_{{q}^{-1}}(a) = R\Delta_{{q}^{-1}}(a)R^{-1},
\hsp \forall\,\,a \in U_{q}(g) \ ,
\label{R1}
\ed
\bn
(\Delta_{{q}^{-1}}\otimes {\rm id})R = R^{13}R^{23} \ , \hsp
({\rm id} \otimes \Delta_{{q^{-1}}})R = R^{13}R^{12},
\label{R2}
\ed
where  $\tilde{\Delta}_{q'}$ is the opposite comultiplication:
$\tilde{\Delta}_{q'}=\sigma\Delta_{q'}$, $\sigma(a \otimes b)=
(-1)^{\deg a \deg b} b\otimes a$ for all homogeneous elements
$a,\, b \in U_{q}(g)$. In (\ref{R2}) we use the standard notation
$R^{12} = \sum a_{i}\otimes b_{i}\otimes {\rm id}$,
$R^{13} = \sum a_{i}\otimes {\rm id} \otimes b_{i}$, $R^{23}=
\sum {\rm id}\otimes a_{i}\otimes b_{i}$  if $R$ has the form
$R=\sum a_{i}\otimes b_{i}$.

Fix some normal ordering in $\underline{\Delta}_{+}$,
and let $e_{\alpha}$ be the corresponding Cartan-Weyl generators
constructed by our procedure.
The following statement holds for any quantized contragredient Lie
(super)algebra of finite growth (see [KT4]).
\begin{theorem}
The equation (\ref{R1}) has a unique (up to a multiplicative constant)
invertible solution in the space $T_{q}(b_{+}\otimes b_{-})$ and
this solution  (for a certain value of the constant) has the form
\bn
R=(\prod_{\alpha \in \underline{\Delta}_{+}}^{\rightarrow}R_{\alpha})
\cdot K \ ,
\label{R3}
\ed
where the order in the product coincides  with  the chosen normal
ordering of $\underline{\Delta}_{+}$ and the elements $R_{\alpha}$ and
$K$ are defined by the formulae:\
\bn
R_{\alpha}= \exp_{q_{\alpha}^{-1}}\left((-1)^{\theta(\alpha)}(q-q^{-1})
(a(\alpha))^{-1}(e_{\alpha}\otimes e_{-\alpha})\right)
\label{R4}
\ed
for any real root $\alpha \, \in \underline{\Delta}_{+}$ and
\bn
R_{n\delta}= \exp((-1)^{\theta(n \delta)}(q-q^{-1})\sum_{i,j}^{mult}
c_{ij}(n)(e^{(i)}_{n\delta} \otimes e^{(j)}_{-n\delta}))
\label{R5}
\ed
for any imaginary root $n\delta\, \in \underline{\Delta}_{+}$ and\
\bn
K=q^{\sum_{i,j} d_{ij} (h_{\alpha_{i}}\otimes h_{\alpha_{j}})} \ ,
\label{R6}
\ed
where $a(\alpha)$ is a factor from the relation (\ref{CW7}) and
$(c_{ij}(n))$ is an inverse to the matrix $(a_{ij}(n))$ with
the elements determined from the relation (\ref{CW5}), and
$d_{ij}$ is an inverse matrix for the symmetrical Cartan matrix
$(a_{ij}^{sym})$ if $(a_{ij}^{sym})$ is not degenerated.
(In the case of degenerated $(a_{ij}^{sym})$ we extend it
up to a non-degenerated matrix $(\tilde{a}_{ij}^{sym})$ and take
an inverse to this extended  matrix (see {\rm [KT1,TK]})).
Moreover the solution (\ref{R3}) is the universal R-matrix, i.e.
it satisfies the equations (\ref{R2}) also.
\end{theorem}

\noindent
The proof of this theorem was given in [KT4] for all quantized
contragredient Lie algebras of finite-dimensional growth.
The explicit formula for the universal R-matrix was obtained in
[Ro,KR,LS] for  the case of the quantized simple Lie algebras
and in [KT1] for the supercase, and in [TK,KT4] for the affine case.

\setcounter{equation}{0}
\section{Twisting of the Hopf structure for $U_{q}(g)$}

In this section we consider at first some general properties of
twisting for an arbitrary Hopf (super)algebra and then return to
the quantum (super)algebra $U_{q}(g)$ again.
\newpage
\noindent
{\large \bf (i) Twisting by Two-Tensor.}

\noindent
Let ${\cal H}_{{\cal A}}:=({\cal A}, \Delta, S, \varepsilon)$ be a
Hopf (super)algebra with a comultiplication $\Delta$, an antipode $S$
and a counit $\epsilon$. Let $F$ be an invertible even element of some
extension $T ({\cal A }\otimes {\cal A})$ of ${\cal A}\otimes {\cal A}$,
such that the formula
\bn
\Delta^{(F)}(a):= F\Delta(a)F^{-1}, \hsp \forall\, a \in {\cal A},
\label{T1}
\ed
determine a new comultiplication, i.e. $\Delta^{(F)}$ satisfies
the coassociativity
\bn
(\Delta^{(F)}\otimes {\rm id})\Delta^{(F)}=
({\rm id}\otimes\Delta^{(F)})\Delta^{(F)}.
\label{T2}
\ed
Then the comultiplication $\Delta^{(F)}$ is called the twisted
coproduct. One can prove the following simple proposition
(see [D1], [R]).

\begin{proposition}
If an invertible even element $F = \sum_{i} f_{i} \otimes f^{i} \in
T ({\cal A }\otimes {\cal A})$ satisfies
the conditions
$$(\varepsilon\otimes {\rm id})F=({\rm id}\otimes\varepsilon)F=1$$
\bn
(F\otimes {\rm id})(\Delta\otimes {\rm id})F =
({\rm id} \otimes F)({\rm id} \otimes \Delta)F ,
\label{T3}
\ed

\noindent
then the element $u:= (({\rm id} \otimes S)F) \circ 1 =
\sum_{i} f_{i} S(f^{i})$ is invertible and the data \
\bn
({\cal A}, \; \Delta^{(F)}, S^{(F)}, \varepsilon)
\label{T4}
\ed
is defined a new Hopf algebra ${\cal H}_{{\cal A}}^{(F)}$, where\
\bn
\Delta^{(F)}(a):= F\Delta(a)F^{-1}, \hsp
S^{(F)}(a):= uS(a)u^{-1}
\label{T5}
\ed
for any $a \in {\cal A}$.
\end{proposition}
(In (\ref{T3}) the comultiplication $\Delta$ acts on components of $F$).

\noindent
The  Hopf (super)algebra ${\cal H}_{\cal A}^{(F)}$ is called the twisted
one by the two-tensor $F$ (or the twisting of the type I).
One should stress that for such twisting the algebraic sector ${\cal A}$
of ${\cal H}_{{\cal A}}$ is not changed but the coalgebraic sector of
${\cal H}_{{\cal A}}$ is changed.
\vspace{4 mm}

\noindent
{\large \bf (ii) Twisting by Automorphism.}

\noindent
We can obtain a twisting of the coalgebraic sector of the Hopf
(super)algebra ${\cal H}_{{\cal A}}$ =
$({\cal A}, \; \Delta, S, \epsilon)$
by using any automorphism in the algebraic sector ${\cal A}$.
Namely, let $\omega: {\cal A} \mapsto {\cal A}$ be an even automorphism
of a linear and multiplicative structure, i.e.
\bn
\hsp \omega(xa+yb) = x \omega a+y \omega b \ , \hsp
\omega (ab)= (\omega a) (\omega b)
\label{T6}
\ed
for any $a, b \in {\cal A}$ and any $x,y \in \bf C$.
The following simple proposition holds.

\begin{proposition}
Let $\Delta^{(\omega)}: {\cal A} \mapsto {\cal A} \otimes {\cal A}$ and
$S^{(\omega)}: {\cal A} \mapsto {\cal A}$ be defined as follows
\bn
\Delta^{(\omega)}(a) := (\omega \otimes \omega)\Delta(\omega^{-1}a) \ ,
\hsp S^{(\omega)}(a) := \omega S (\omega ^{-1}a) \ ,
\label{T7}
\ed
for any $a \in {\cal A}$. Then ${\cal H}_{\cal A}^{(\omega)}:=
({\cal A} \ , \Delta^{(\omega)}, S^{(\omega)}, \varepsilon)$
is a Hopf (super)algebra isomorphic to
${\cal H}_{{\cal A}}= ({\cal A}, \; \Delta, S, \epsilon)$.
If ${\cal H}_{{\cal A}}$ is quasitriangular with an universal
R-matrix $R$ then ${\cal H}_{{\cal A}}^{(\omega)}$ is also
quasitriangular with the universal R-matrix $R^{(\omega)}$:
\bn
R^{(\omega)} = (\omega \otimes \omega)R \ .
\label{T8}
\ed
\end{proposition}
The Hopf (super)algebra ${\cal H}_{{\cal A}}^{(\omega)}$ is called the
twisted Hopf (super)algebra by the automorphism $\omega$
(or the twisting of the type II). Note that the algebra structure for
the twisting of the type II does not change also.
\vspace{4 mm}

\noindent
{\large \bf(iii) Twisting for $U_{q}(g)$ by some Factors of
the Universal R-matrix.}

\noindent
Now we want to show how some factors of the universal R-matrix generate
a family of twistings for $U_{q}(g)$.

For a fixed normal ordering in $\underline{\Delta}_{+}$ and for any
$\gamma \, \in \underline{\Delta}_{+}$ we put
\bn
F_{\gamma} := \prod_{\alpha < \gamma} R^{21}_{\alpha}
\label{T9}
\ed
and
\bn
F'_{\gamma} := (\prod_{\gamma < \alpha} R^{}_{\alpha})K
\label{T10}
\ed
where $R_{\alpha}$ are the factors of the universal R-matrix (\ref{R3})
and the product in (\ref{T9}) (and (\ref{T10})) is taken over all roots
$\alpha$ which are less (and more) $\gamma$ in a sense of the normal
ordering. The following theorem is valid.
\begin{theorem}
For any roots $\gamma \in \D$ the two sets of the data \
\bn
(U_{q}(g) \ , \Delta_{{q}^{-1}}^{(F_{\gamma})},
S_{{q}^{-1}}^{(F_{\gamma})} , \varepsilon) \ , \;\;\;and \;\;\;
(U_{q}(g) \ , \Delta_{q^{-1}}^{(F'_{\gamma})} \ ,
S_{{q^{-1}}}^{(F'_{\gamma})} , \varepsilon)
\label{T11}
\ed
are defined two Hopf algebras, where $\Delta_{{q}^{-1}}^{(F_{\gamma})}$,
$S_{{q}^{-1}}^{(F_{\gamma})}$, and
$\Delta_{{q}^{-1}}^{(F'_{\gamma})}$, $S_{{q}^{-1}}^{(F'_{\gamma})}$
are determined by the formulas
\bn
\Delta_{{q}^{-1}}^{(F_{\gamma})}(a):=
F_{\gamma}^{-1}\Delta_{{q}^{-1}}(a)F_{\gamma} \ , \hsp
S_{{q}^{-1}}^{(F_{\gamma})}(a) =
u_{\gamma}^{-1} S_{{q}^{-1}}(a)u_{\gamma},
\label{T12}
\ed
and
\bn
\Delta_{{q}^{-1}}^{(F'_{\gamma})}(a):=
F'_{\gamma}\Delta_{{q}^{-1}}F_{\gamma}'^{-1}, \hsp
S_{{q}^{-1}}^{(F'_{\gamma})}(a) =
u'_{\gamma}S_{{q}^{-1}}(a)u_{\gamma}'^{-1}.
\label{T13}
\ed
for any $a \in U_q (g)$. (Here in (\ref{T12}) and (\ref{T13})
the elements $u_{\gamma}$ and $u'_{\gamma}$ are determined by
the formulas similar to $"u"$ in the Proposition 5.1).
\end{theorem}

\noindent
{\it Proof}. Using the Proposition 8.3 of [KT1], which is valid
indeed for any quantized contragredient Lie (super)algebra of finite
growth, we prove at first that $\Delta_{{q}^{-1}}^{(F_{\gamma})}$ and
$\Delta_{{q}^{-1}}^{(F'_{\gamma})}$ satisfy the coassociativity
(\ref{T3}) after that we apply the Proposition 5.1.

Now we would like to show that for the case of a quantized simple
finite-dimensional Lie algebra $g$ the twisting by the two-tensor
$F_{\gamma}$ coincides with the twisting by the Lusztig automorphism of
$U_q(g)$. First we remind that there is one-to-one correspondence
between the set of all normal orderings of $\Delta_+$ and the set of
reduced decompositions of the longest element $w_{o}$ of the Weyl
group $W(g)$. Namely, the following proposition is valid (see [Z]).
\begin{proposition}
Let $s_{{\alpha}_i}$, $(i=1,2, \ldots, r)$, be elementary
reflections of $W(g)$, corresponding to the simple roots
$\alpha_i$ and let
\bn
\gamma_{1},\; \gamma_{2},\; \gamma_{3},\; \ldots,\; \gamma_{n},\;
\ldots,\; \gamma_{N}
\label{T14}
\ed
be a fixed normal ordering in $\Delta_+$, then all roots of the chain\
\[
\alpha_{i_1}=\gamma_{1},\;\; \alpha_{i_2}=
s^{-1}_{\alpha_{i_1}}(\gamma_{2}), \;\;
\alpha_{i_3}=s^{-1}_{\alpha_{i_2}}s^{-1}_{\alpha_{i_1}}(\gamma_{3}),\;\;
\ldots\;\;,
\]
\bn
\alpha_{i_n}= s^{-1}_{\alpha_{i_{n-1}}}\cdots s^{-1}_{\alpha_{i_2}}
s^{-1}_{\alpha_{i_1}}(\gamma_{n}), \;\;
\ldots\;\; , \alpha_{i_N}= s^{-1}_{\alpha_{i_{N-1}}}\cdots
s^{-1}_{\alpha_{i_2}}s^{-1}_{\alpha_{i_1}}(\gamma_{N})
\label{T15}
\ed
are simple, and  $w_{o}=s_{\alpha_{i_1}} s_{\alpha_{i_2}}
\cdots s_{\alpha_{i_N}}$ is a reduced decomposition of $w_{o}$.

\noindent
On the contrary, for any reduced decomposition of $w_{o}$,
$w_{o}=s_{\alpha_{i_1}} s_{\alpha_{i_2}} \cdots s_{\alpha_{i_N}}$,
the sequence
\bn
\alpha_{i_1},\;\; s_{\alpha_{i_1}}(\alpha_{i_2}),\;\;
s_{\alpha_{i_1}}s_{\alpha_{i_2}}(\alpha_{i_3}),\;\;\ldots\;\;,
s_{\alpha_{i_1}}s_{\alpha_{i_2}} \cdots s_{\alpha_{i_{N-1}}}
(\alpha_{i_N})
\label{T16}
\ed
is a normal ordering in $\Delta_{+}$.
\end{proposition}

\noindent
(It should be noted that there are identical simple roots in
(\ref{T15}) because $r < N$).

\noindent
For every root $\gamma_n$, $(n=1,2, \ldots, N)$, of the normal ordering
(\ref{T14}) we put into correspondence the element $w_{\gamma_n}$ of
the Weyl group $W(g)$, which is the initial segment of the corresponding
reduced decomposition of $w_{o}$, i. e.
\bn
w_{\gamma_n}:=s_{\alpha_{i_1}}s_{\alpha_{i_2}}\cdots
s_{\alpha_{i_{n-1}}} \ ,
\hsp (w_{\gamma_{n}}(\alpha_{i_n}) = \gamma_n) \ .
\label{T17}
\ed

Following [L], \cite{KacdeConchini}, we define an action of
a braid group $\hat{W}(g)$ in $U_{q}(g)$ by means of the Lusztig
automorphisms (see (\ref{Q3}))
\[
{\hat{s}_{\alpha_{i}}}(k_{\alpha_i}^{\pm 1})=
k_{\alpha_i}^{\mp 1} \ , \hsp
{\hat{s}_{\alpha_{i}}}(k_{\alpha_j}^{\pm 1})=k_{\alpha_j}^{\pm 1}
k_{\alpha_i}^{\pm n_{ij}} \ , \;\;\;\; (i\neq j) \ ,
\]
\bn
{\hat{s}_{\alpha_{i}}}(e_{\alpha_{i}})=
-e_{-\alpha_{i}} k_{\alpha_{i}} \ , \hsp
{\hat{s}_{\alpha_{i}}}(e_{-\alpha_{i}})=
-k_{\alpha_i}^{-1}e_{\alpha_{i}} \ ,
\label{T18}
\ed
\[
{\hat{s}_{\alpha_{i}}}(e_{\pm \alpha_{j}})=
N_{ij}^{-\frac{1}{2}}(\tilde{\rm ad}_{q}\epma{i})^{n_{ij}}\epma{j} \ ,
\;\;\;\; (i \neq j) \ ,
\]
\noindent
where the positive integer $n_{ij}$ are determined by the formula
(\ref{Q5}) (the last line for this case); the normalizing
factors $N_{ij}$ can be determined from results of the work [KT1],
and they have the form
\bn
N_{ij} = q^{(\alpha_{i},\alpha_{j})}
([(\alpha_{i}, \alpha_{j})]_{q})^{n_{ij}} \prod_{s=1}^{n_{ij}-1}
\frac{[(s\alpha_{i}+\alpha_{j}, s\alpha_{i}+\alpha_{j})]_{q}}
{[((s-1)\alpha_{i}+\alpha_{j},s\alpha_{i}+\alpha_{j})]_{q}} \ .
\label{T19}
\ed
For the element $w_{\gamma_{n}}$ (\ref{T17}) we put
\bn
\hat{w}_{\gamma_n}:= \hat{s}_{\alpha_{i_1}} \hat{s}_{\alpha_{i_2}}\cdots
\hat{s}_{\alpha_{i_{n-1}}} \ .
\label{T20}
\ed

\noindent
The following statement (known in quantum group folklore) holds.
\begin{proposition}
For any normal ordering (\ref{T14}) of any simple Lie algebra $g$ the
twisting of the Hopf algebra structure of the quantum algebra $U_{q}(g)$
by the two-tensor $F_{\gamma_{n}}$ (see (\ref{T9}), (\ref{T12}))
coincides with the twisting by the Lusztig automorphism
$\hat{w}_{\gamma_{n}}$
(\ref{T18}), i.e. $\Delta_{q^{-1}}^{(F_{\gamma_{n}})}=
\Delta_{q^{-1}}^{(\hat{w}_{\gamma_{n}})}, \,
S_{q^{-1}}^{(F_{\gamma_{n}})}= S_{q^{-1}}^{(\hat{w}_{\gamma_{n}})}$ .
\end{proposition}
{\it Proof}. We prove the statement for a root $\gamma_{n}$ of
(\ref{T14}) by induction on $n$.
For  $n = 2$ we have that $\gamma_{1}=\alpha_{1}$ is a simple root,
$F_{\gamma_{2}} = \exp_{q_{\alpha_1}^{-1}}((q-q^{-1})(e_{-\alpha_{1}}
\otimes e_{\alpha_{1}}))$, and $\hat{w}_{\gamma_{2}}$ is the Lusztig
automorphism, corresponding to the simple reflection $s_{\alpha_{1}}$.
It is not difficult to verify that \
\bn
\Delta_{q^{-1}}^{(F_{\gamma_{2}})}(e_{\alpha_1})=
\Delta_{q^{-1}}^{(\hat{w}_{\gamma_{2}})}(e_{\alpha_1})=
e_{\alpha_1}\otimes 1 + k_{\alpha_1}\otimes e_{\alpha_1} \ ,
\label{T21}
\ed
\bn
\Delta_{q^{-1}}^{(F_{\gamma_{2}})}(e_{-\alpha_1})=
\Delta_{q^{-1}}^{(\hat{w}_{\gamma_{2}})}(e_{-\alpha_1})=
e_{-\alpha_1}\otimes k^{-1}_{\alpha_1} + 1\otimes e_{-\alpha_1} \ ,
\label{T22}
\ed
and also
\bn
\Delta_{q^{-1}}^{(F_{\gamma_{2}})}(e_{\alpha_i})=
\Delta_{q^{-1}}^{(\hat{w}_{\gamma_{2}})}(e_{\alpha_i})=
e_{\alpha_i}\otimes 1 + k^{-1}_{\alpha_i}\otimes e_{\alpha_i} \ ,
\label{T23}
\ed
\bn
\Delta_{q^{-1}}^{(F_{\gamma_{2}})}(e_{-\alpha_i})=
\Delta_{q^{-1}}^{(\hat{w}_{\gamma_{2}})}(e_{-\alpha_i})=
e_{-\alpha_i}\otimes k_{\alpha_i} + 1\otimes e_{-\alpha_i} \ ,
\label{T24}
\ed
for any simple root $\alpha_{i} \neq \alpha_1$ such that
$(\alpha_1,\alpha_i)=0$.
Now let $\alpha_{i} \neq \alpha_1$ be a simple root such that
$(\alpha_1,\alpha_i)\neq 0$
then $\beta_i:= \hat{w}_{\alpha_1}(\alpha_i)$
is a positive root and one can choose a normal ordering
\bn
\gamma'_{1},\; \gamma'_{2},\; \gamma'_{3},\; \ldots,\; \gamma'_{N}
\label{T25}
\ed
such that $\gamma'_{1}=\alpha_1$, $\gamma'_{2}=\beta_i$. Using the
Proposition 8.3 from [KT1] we have
\bn
\Delta_{q^{-1}}^{(F_{\gamma_{2}})}(e_{\beta_i})=
\Delta_{q^{-1}}^{(\hat{w}_{\gamma_{2}})}(e_{\beta_i})=
e_{\beta_i}\otimes 1 + k^{-1}_{\beta_i}\otimes e_{\beta_i} \ ,
\label{T26}
\ed
\bn
\Delta_{q^{-1}}^{(F_{\gamma_{2}})}(e_{-\beta_i})=
\Delta_{q^{-1}}^{(\hat{w}_{\gamma_{2}})}(e_{-\beta_i})=
e_{-\beta_i}\otimes k_{\beta_i} +  1\otimes e_{-\beta_i} \ .
\label{T27}
\ed
The equations (\ref{T21})-(\ref{T24}) and (\ref{T26}), (\ref{T27})
are sufficient to conclude that
\bn
\Delta_{q^{-1}}^{(F_{\gamma_2})} =
\Delta_{q^{-1}}^{(\hat{w}_{\gamma_2})} \ .
\label{T28}
\ed
For $n > 2$ the statement follows immediately from the multiplicative
structure of $\hat{w}_{\gamma_n}$, taking in mind, that
\bn
F_{\gamma_n} = F_{\gamma_{n-1}}R^{21}_{\gamma_{n-1}}=
F_{\gamma_{n-1}}(\hat{w}_{\gamma_{n-1}}\otimes \hat{w}_{\gamma_{n-1}})
\exp_{q_{\alpha_{n-1}}^{-1}}((q - q^{-1})(e_{-\alpha_{n-1}}
\otimes e_{\alpha_{n-1}}))
\label{T29}
\ed
in the notations of (\ref{T17}).

\noindent
Since an antipode is determined uniquely by a coproduct, therefore
$S_{q^{-1}}^{(F_{\gamma_{n}})}= S_{q^{-1}}^{(\hat{w}_{\gamma_{n}})}$.
The proof is finished.

An analog of the Proposition 5.4 is also valid for the quantized
superalgebras. In this case the Lusztig automorphisms should be
considered as isomorphisms between different quantized superalgebras.
Let us consider this in detail.

\noindent
Let $g(A,\Upsilon)$ and $g(A',\Upsilon')$ be two isomorphic
finite-dimensional contragredient Lie superalgebras.
We consider non-trivial case when the superalgebras have non-equivalent
Cartan matrices $A$ and $A'$, i. e. $A'\neq BDAB^{-1}$ for any
nonsingular matrix $B$ and any diagonal matrix $D$.
Such superalgebras have the same reduced system of all roots and
different reduced system of positive roots.

\noindent
Let $\Pi:= \{\alpha_{1},\alpha_{2}, \ldots , \alpha_{r}\}$
and $\Pi':= \{\alpha_{1}',\alpha_{2}', \ldots , \alpha_{r}'\}$ be
systems of simple roots for  $g(A,\Upsilon)$ and $g(A',\Upsilon')$
correspondingly.
Following Serganova (see [LSS]) one can define an "elementary
reflection" $s_{\alpha_i}$ for any $\alpha_i \in \Pi$  as follows
\bn
s_{\alpha_i}({\alpha}_i):=\alpha_i'= -\alpha_i \ , \hsp
s_{\alpha_i}(\alpha_j):=\alpha_j'=
\alpha_j-n_{ij}\alpha_i,\hsp (i\neq j) \ .
\label{T30}
\ed

\noindent
The following theorem is valid (see [LSS]).
\begin{theorem} {\rm (V. Serganova)}.
{\rm (i)} Let $\Pi$ be a system of simple roots. Then the set
$s_{\alpha_i}(\Pi)$ may be considered as the system of simple roots,
moreover $s_{\alpha_i}(\Pi)$ is not equivalent to $\Pi$ iff the root
$\alpha_i$ is grey.

\noindent
{\rm (ii)} For any two isomorphic superalgebras  $g(A,\Upsilon)$ and
$g(A',\Upsilon')$ with non-equivalent Cartan matrices $A$ and $A'$
there exist a sequence of simple root systems $\Pi_1,  \Pi_2,
\ldots, \Pi_n$ and roots $\alpha^{(i)} \in \Pi_i$ such that
$s_{\alpha^{(i)}}(\Pi_i)= \Pi_{i+1}$, and moreover $\Pi_1=\Pi$,
$\Pi_n=\Pi'$.
\end{theorem}

\noindent
{\it Remark}. It should be noted that in a general case the system of
all roots $\Delta$ is not invariant with respect to
the "elementary reflection" $s_{\alpha_i}$, i. e.
$s_{\alpha_i}(\Delta)$ is a root system which should
not coincide with $\Delta$ (see, for example, the root system of the
superalgebra $D(2,1;\alpha)$ [K2]). (This is why the words "elementary
reflection" are putted in quotation marks). Therefore it is better to
consider the "elementary reflection" as change of variables or as a map
from one linear space to another. An accurate formulation leads to the
notation of Weyl grouppoid instead of the Weyl group.

\noindent
Let simple root systems $\Pi$ and $\Pi'$ are connected by
one "elementary reflection" $s_{\alpha_{i}}$ only, i.e.
$s_{\alpha_{i}}(\Pi) = \Pi'$, and
let  $\{e_{\pm \alpha_{j}},\; k_{\alpha_{j}}^{\pm 1} \}$ and
$\{e_{\pm \alpha_{j}'}', \; k_{\alpha_{j}'}'^{\pm 1} \}$,
$(j=1,2,\ldots,r)$, be the Chevalley generators of the quantum algebras
$U_{q}(g(A,\Upsilon))$ and $U_{q}(g(A',\Upsilon'))$ correspondingly.
We define the Lusztig isomorphism $\hat{s}_{\alpha_i}:
U_q(g(A',\Upsilon'))\mapsto U_q(g(A,\Upsilon))$ by the relations
analogous to (\ref{T18}):
\[
{\hat{s}_{\alpha_{i}}}(k_{\alpha_i'}'^{\pm 1})=k_{\alpha_i}^{\mp 1} ,
\hsp {\hat{s}_{\alpha_{i}}}(k_{\alpha_j'}'^{\pm 1})=k_{\alpha_j}^{\pm 1}
k_{\alpha_i}^{\pm n_{ij}} , \;\;\;\; (i\neq j) \ ,
\]
\bn
{\hat{s}_{\alpha_{i}}}(e_{\alpha_{i}'}')=
-e_{-\alpha_{i}} k_{\alpha_{i}} \ ,\hsp
{\hat{s}_{\alpha_{i}}}(e_{-\alpha_{i}'}')=
-k_{\alpha_i}^{-1}e_{\alpha_{i}} \ ,
\label{T31}
\ed
\[
{\hat{s}_{\alpha_{i}}}(e_{\pm\alpha_{j}'}')=
((-1)^{\theta(n_{ij}\alpha_i) \theta (\alpha_j)}N_{ij})^{-\frac{1}{2}}
(\tilde{\rm ad}_{q}\epma{i})^{n_{ij}}\epma{j} \ ,\;\;\;\; (i \neq j) \ ,
\]
where the normalizing factors  $N_{ij}$ are given by the formula
(\ref{T19}).

\noindent
If $\Delta_{q^{-1}}$ and $\Delta_{q^{-1}}'$ are the standard
comultiplications of $U_{q}(g(A,\Upsilon))$ and $U_{q}(g(A',\Upsilon'))$
(see (\ref{Q11})-(\ref{Q13})) then just as in the even case using the
Proposition 8.3 from [KT1] we have
\bn
(\hat{s}_{\alpha_{i}} \otimes \hat{s}_{\alpha_{i}}) \Delta_{q^{-1}}'(a')
= (\hat{s}_{\alpha_{i}} \otimes \hat{s}_{\alpha_{i}})
\Delta'_{q^{-1}}(\hat{s}_{\alpha_{i}}^{-1}a) =
F^{-1}_{\alpha_{i}}\Delta_{q^{-1}}(a)F_{\alpha_{i}}
\label{T32}
\ed
for any $a \in U_{q}(g(A,\Upsilon))$, where
\bn
F_{\alpha_{i}}=R^{21}_{\alpha_{i}}:=
\exp_{q_{\alpha_{i}}^{-1}}((q-q^{-1})
(e_{-\alpha_{i}} \otimes e_{\alpha_{i}})) \ .
\label{T33}
\ed
Now let $\Pi$ and $\Pi'$ be  arbitrary non-equivalent simple root
systems of the isomorphic superalgebras $g(A,\Upsilon)$ and
$g(A',\Upsilon')$ correspondingly. Then according to the
Serganova's theorem there is a chain
$w:=s_{\alpha^{(1)}}s_{\alpha^{(2)}}\cdots s_{\alpha^{(n)}}$ of the
elementary reflections (\ref{T30}), such that
$\alpha_{i}'= w(\alpha_{i})$, $(i=1,2,\ldots,r)$.
(Here we do not distinguish simple root systems
which differ by an enumeration of roots). We define the Lusztig
isomorphism $\hat{w}: U_q(g(A',\Upsilon'))\mapsto U_q(g(A,\Upsilon))$
as  follows
\bn
\hat{w}:=\hat{s}_{\alpha^{(1)}} \hat{s}_{\alpha^{(2)}} \cdots
\hat{s}_{\alpha^{(n)}} \ .
\label{T34}
\ed
The relation (\ref{T32}) turns to
\bn
(\hat{w} \otimes \hat{w}) \Delta_{q^{-1}}'(a') =
(\hat{w} \otimes \hat{w}) \Delta'_{q^{-1}} (\hat{w}^{-1}a) =
F^{-1}_{w}\Delta_{q^{-1}}(a)F_{w}
\label{T35}
\ed
for any $a \in U_{q}(g(A,\Upsilon))$, where the twisting two-tensor
$F_{w}$ is defined by the formula
\bn
F_{w}=F_{\alpha^{(1)}}((\hat{s}_{\alpha^{(1)}} \otimes
\hat{s}_{\alpha^{(1)}})F_{\alpha^{(2)}})\cdots ((\hat{s}_{\alpha^{(1)}}
\cdots \hat{s}_{\alpha^{(n-1)}}\otimes \hat{s}_{\alpha^{(1)}}
\cdots\hat{s}_{\alpha^{(n-1)}})F_{\alpha^{(n)}}) \ .
\label{T36}
\ed
It should be noted that the factors $(\hat{s}_{\alpha^{(1)}}
\cdots\hat{s}_{\alpha^{(l-1)}}\otimes\hat{s}_{\alpha^{(1)}}
\cdots\hat{s}_{\alpha^{(l-1)}})F_{\alpha^{(l)}}$, $l=1,2,\ldots,n$,
belong to $T_{q}(g(A,\Upsilon) \otimes g(A,\Upsilon))$ and are composed
from factors of the universal R-matrix of $U_{q}(g(A,\Upsilon))$.

We can summarize our considerations as the theorem.
\begin{theorem}
Let $g(A,\Upsilon)$ and $g(A',{\Upsilon}')$ be two isomorphic
finite-dimensional contragredient Lie superalgebras. Then there exists
an isomorphism of algebras, $\hat{w}: U_{q}(g(A',\Upsilon')) \mapsto
U_{q}(g(A,\Upsilon))$, such that a comultiplication
$\Delta_{q^{-1}}^{(\omega)}$ of $\;$ $U_{q}(g(A,\Upsilon))$ induced from
$\Delta_{q^{-1}}'$ of $\;$  $U_{q}(g(A',\Upsilon'))$ by the isomorphism
$\omega$ differs from the initial comultiplication $\Delta_{q^{-1}}$
with a twisting by some factors of the universal R-matrix for
$U_{q}(g(A,\Upsilon))$.
\end{theorem}

\setcounter{equation}{0}
\section{Drinfeld's realization of quantum affine algebras}

In this and last sections we consider quantized
nontwisted affine algebras only.

Let $\hat{g}$ be a nontwisted affine Lie algebra  and
$\Pi =\{ \alpha_{0}, \alpha_{1},\ldots, \alpha_{r}\}$ be a system
of simple roots  for $\hat{g}$. We assume that the roots
$\Pi_0  =\{ \alpha_{1}, \alpha_{2},\ldots, \alpha_{r}\}$ generate
the system $\Delta_+ (g)$ of positive roots of the corresponding
finite-dimensional Lie algebra $g$.

In the paper "A new realization of Yangians and quantized
affine algebras" [D2] V.G. Drinfeld suggested another realization of
the nontwisted affine algebra $U_q(\hat{g})$. In this description
the algebra $U_q(\hat{g})$ is generated by the elements:
\bn
k_c,\,\,\,\chi_{i,l},\,\,\,\xi_{i,l}^\pm \ , \hsp ({\rm for}\;\;
i=1,2,\ldots ,r;\;\;\; l \in {\bf Z}) \ ,
\label{D1}
\ed
with the defining  relations (we modify them a little for technical
 convenience):
\bn
[k_c,{\rm  everything}]=0 \ , \hsp \chi_{i,0}\xi_{j,m}^\pm =
q^{\pm (\alpha_i,\alpha_j)}\xi_{j,m}^\pm\chi_{i,0} \ ,
\label{D2}
\ed
\bn
[\chi_{i,l},\chi_{j,m}] = \delta_{l,-m} a_{ij}(l)
\,\frac{k_c^{l} - k_c^{-l}}{q - q^{-1}} \ ,
\label{D3}
\ed
\bn
[\chi_{i,l},\xi_{j,m}^\pm]=
\pm a_{ij}(l) \,\xi_{j,l+m}^\pm k_c^{(-l\pm |l|)/2} \ ,
\label{D4}
\ed
\bn
\xi_{i,l+1}^\pm \xi_{j,m}^\pm -q^{\mp(\alpha_{i},\alpha_{j})}
\xi_{j,m}^\pm\xi_{i,l+1}^\pm =q^{\mp(\alpha_{i},\alpha_{j})}
\xi_{i,l}^\pm\xi_{j,m+1}^\pm -\xi_{j,m+1}^\pm\xi_{i,l}^\pm \ ,
\label{D5}
\ed
\bn
[\xi_{i,l}^+,\xi_{j,m}^-]=\delta_{i,j} \,
\frac{\phi_{i,l+m}k_c^{l}-\psi_{i,l+m} k_c^{m}}{q - q^{-1}} \ ,
\label{D6}
\ed
\bn
{\rm Sym} \left(\sum_{s=o}^{n_{ij}'}(-1)^{s}
C_{n_{ij}'}^{s}(q^{(\alpha_{i},
\alpha_{j})})\xi_{i,l_{1}}^\pm  \cdots \xi_{i,l_{s}}^\pm \xi_{j,m}^\pm
\xi_{i,l_{s+1}}^\pm \cdots \xi_{i,l_{n_{ij}'}}^\pm  \right)= 0\;\;\;\;\;
{\rm for}\;\; i\neq j \,
\label{D7} \ ,
\ed
where
\bn
a_{ij}(l) =
\frac{q^{l(\alpha_{i},\alpha_{j})}-q^{-l(\alpha_{i},\alpha_{j})}}
{l(q - q^{-1})} \ ,
\label{D8}
\ed
the elements $\phi_{i,p}$, $\psi_{i,p}$ are defined from the relations:
\bn
\sum_{p} \phi_{i,p}u^{-p} = \chi_{i,0}\exp ((q^{-1}-q)
\sum_{p<0}\chi_{i,p}u^{-p}) \ ,
\label{D9}
\ed
\bn
\sum_{p} \psi_{i,p}u^{-p} =  \chi_{i,0}^{-1}\exp ((q-q^{-1})
\sum_{p>0}\chi_{i,p}u^{-p}) \ ,
\label{D10}
\ed
the q-binomial coefficients $C_{n}^{s}(q)$ are determined by the formula
\bn
C_{n}^{s}(q)=\frac{[n]_{q}!}{[s]_{q}![n-s]_{q}!} \ ,
\label{D11}
\ed
the symbol ``Sym'' in (\ref{D7}) denotes a symmetrization on $l_1$,
$l_{2}$,\ldots, $l_{n_{ij}}$, and $n_{ij}':= n_{ij} + 1$.

\noindent
It should be noted that the matrix $(a_{ij}(l))$ with the elements
(\ref{D8}) may be considered as a q-analog of the "level $l$'' for
the matrix Cartan $(a^{sym}_{ij})$.

\noindent
Drinfeld has shown how to express the Chevalley generators
$e_{\alpha_{i}}$, $h_{\alpha_{i}}$  in terms of $\chi^{\pm}_{i,0}$
and $\xi^{\pm}_{i,k}$, $k=0,\pm 1$, (see [D2]).
He suggested also another formulas of the comultiplication
for $U_{q}(g)$, which originate in a quantization of the
corresponding bialgebra structure [D3] (different from usual one),:
 \bn
\Delta^{(D)}(k_c)=k_c\otimes k_c \ , \hsp
\Delta^{(D)}(\chi_{i,0})=\chi_{i,0}\otimes \chi_{i,0},
\label{D12}
\ed
\bn
\Delta^{(D)}(\chi_{i,l})=\chi_{i,l}\otimes 1 +
k_c^{-l}\otimes \chi_{i,l}\ , \;\;\;
\Delta^{(D)}(\chi_{i,-l})=\chi_{i,-l}\otimes k_{\delta}^{} +
1\otimes \chi_{i,-l}
\label{D13}
\ed
for $l > 0$, and
\bn
\Delta^{(D)}(\xi_{i,l}^+) =\xi_{i,l}^+\otimes 1 +\sum_{m\geq 0}
k_c^m \phi_{i,m}\otimes \xi_{i,l+m}^+ \ ,
\label{D14}
\ed
\bn
\Delta^{(D)}(\xi_{i,l}^-) =1\otimes\xi_{i,l}^- + \sum_{m\geq 0}
\xi_{i,l-m}^-\otimes \psi_{i,m}k_c^m \ ,
\label{D15}
\ed
for any $l \in {\bf Z}$.

Now we want to show how the generators
$k_c$, $\chi_{i,l}$, $\xi_{i,l}^\pm$
can be expressed via the Cartan-Weyl generators constructed
by our procedure, and in the Section 8 we show how to obtain
the  formulas (\ref{D12})-(\ref{D15}) by a twisting of the standard
comultiplication (\ref{Q12})-(\ref{Q15}) using some factor of the
universal R-matrix.

\setcounter{equation}{0}
\section{Connection of the Drinfeld's realization with the Cartan-Weyl
basis for $U_{q}(\hat{g})$}

We fix some special normal ordering in $\Delta_{+}(\hat{g}) :=
\Delta_{+}$, which satisfies the following additional constraint:
\bn
l\delta + \alpha_{i} <(m+1)\delta < (n+1)\delta -\alpha_{j}
\label{C1}
\ed
for any simple roots $\alpha_{i},\alpha_{j}\in \Pi_0$,
and $l,\,m, \,n \geq 0$.  Here $\delta$ is a minimal positive imaginary
root.  For given normal ordering we apply our procedure for
the construction of the Cartan-Weyl generators for $U_{q}(\hat{g})$.
Furthermore we put
\bn
{e_{\delta}^{(i)}}=[e_{{\alpha}_{i}}, e_{\delta -{\alpha}_{i}}]_{q} \ ,
\label{C2}
\ed
\bn
e_{n\delta + \alpha_{i}}=(-1)^n([(\alpha_i,\alpha_i)]_q)^{-n}
(\tilde{\rm ad} \; e_{\delta}^{(i)})^{n}e_{{\alpha}_{i}} \ ,
\label{C3}
\ed
\bn
e_{(n+1)\delta -\alpha_{i}}=([(\alpha_i,\alpha_i)]_q)^{-n}
(\tilde{\rm ad}\; e_{\delta}^{(i)})^{n}
e_{\delta -\alpha_{i}}  \ ,
\label{C4}
\ed
\bn
e_{(n+1)\delta}'^{(i)}=
[e_{n\delta + \alpha_{i}}, e_{\delta - \alpha_{i}}]_{q} \ ,
\label{C5}
\ed
(for $n > 0$), where $(\tilde{\rm ad}\,x)y=[x,y]$ is a usual
commutator. The imaginary root vectors $e'^{(i)}_{\pm n \delta}$ do
not satisfy the relation (\ref{CW5}). We introduce new vectors
$e^{(i)}_{\pm n \delta}$ by the following (Schur) relations:
\bn
e_{n\delta}'^{(i)}=\sum_{p_{1}+2p_{2}+\ldots +np_{n}=n}
\frac{{(q - q^{-1})}^{\sum p_{i}-1}}{p_{1}!\cdots
p_{n}!}{(e_{\delta}^{(i)})}^{p_1}\cdots {(e_{n\delta}^{(i)})}^{p_n}.
\label{C6}
\ed
In terms of the generating functions
\bn
E'_{i}(z)= (q-q^{-1})\sum_{m\geq 1}^{} e_{m\delta}'^{(i)}z^m
\label{C7}
\ed
and
\bn
E_{i}(z)= (q-q^{-1})\sum_{m\geq 1}^{} e_{m\delta}^{(i)}z^m
\label{C8}
\ed
the relation (\ref{C6}) may be rewritten in the form
\bn
E'_{i}(z)= -1 +\exp E_{i}(z)
\label{C9}
\ed
or
\bn
E_{i}(z)=  \ln (1 + E'_{i}(z)) \ .
\label{C10}
\ed
{}From this we have the inverse formula to (\ref{C6})
\bn
e_{n\delta}^{(i)}=\sum_{p_{1}+2p_{2}+\ldots +np_{n}=n}
\frac{{(q^{-1} - q)}^{\sum p_{i}-1} (\sum_{i=1}^{n}p_{i} - 1)!}
{p_{1}!\cdots p_{n}!}{(e_{\delta}'^{(i)})}^{p_1}
\cdots {(e_{n\delta}'^{(i)})}^{p_n}.
\label{C11}
\ed
The rest of the real root vectors we construct in accordance with
the Algorithm 3.1 using the root vectors $e_{n\delta + \alpha_i}$,
$e_{(n+1)\delta - \alpha_i}$, $e^{(i)}_{(n+1)\delta}$,
($i = 1,2, \ldots, r$; $n \in {\bf Z_{+}}$). The root vectors of
negative roots are obtained by the Cartan involution $(^*)$:
\bn
e_{-\gamma}=(e_{\gamma})^{*}
\label{C12}
\ed
for $\gamma  \in \Delta (\hat{g})$.

\noindent
Using the explicit relations
(\ref{C2})-(\ref{C5}), (\ref{C11}) and (\ref{C12}) we can prove the
following theorem
which states the connection between the Cartan-Weyl
and Drinfeld's generators for the quantum nontwisted affine algebra
$U_q(\hat g)$.
\begin{theorem}
Let some function $\pi$: $\{ \alpha_1, \alpha_2\,\ldots ,
\alpha_r \}\mapsto \{ 0,1\}$ be chosen  such  that
$\pi (\alpha_i)$ $\neq$ $\pi (\alpha_j)$ if $(\alpha_i,\alpha_j)\neq 0$
and let  the root vectors $\hat{e}_{\pm\gamma}$ and
$\check{e}_{\pm\gamma}$
of the real roots $\gamma \in \Delta _{+}(\hat{g})$
be the circular Cartan-Weyl generators (\ref{CW12}), (\ref{CW13}) and
$e_{n\delta}^{(i)}$ be imaginary root vectors of $U_q(\hat{g})$.
Then the elements
\bn
k_c:=k_\delta \ , \hsp \chi_{i,0}:=k_{\alpha_i} \ , \hsp
\chi_{i,n}:=(-1)^{n\pi(\alpha_i)}e_{n\delta}^{(i)},
\label{C13}
\ed
\bn
\xi^+_{i,n}=(-1)^{n\pi(\alpha_i)}\hat{e}_{n\delta + \alpha_i} \ , \hsp
\xi^-_{i,n}=(-1)^{n\pi(\alpha_i)}\check{e}_{n\delta - \alpha_i} \ ,
\label{C14}
\ed
for $n \in {\bf Z}$, and
$$
\phi_{i,0}=k_{\alpha_i} \ , \hsp
\phi_{i,-n}=(q^{-1}-q)k_{\alpha_i}e_{-n\delta}'^{(i)} \ ,
$$
\bn
\psi_{i,0}=k_{\alpha_i}^{-1} \ ,\hsp
\psi_{i,n}=(q-q^{-1})k_{\alpha_i}^{-1}e_{n\delta}'^{(i)} \ ,
\label{C15}
\ed
for  $n > 0$ satisfy the relations (\ref{D2})-(\ref{D7}),
i.e. the elements (\ref{C13})-(\ref{C15}) are the generators of
the Drinfeld's realization of $U_q(\hat{g})$.
\end{theorem}

\noindent
{\it Remark}. In terms of the Cartan-Weyl generators the relations
(\ref{D2})-(\ref{D7}) can be interpreted as follows:

\noindent
(i) The "Serre" relations (\ref{D7}) are equivalent to the following
corollary of the Proposition 3.1:
\bn
[e_\alpha ,e_\beta ]_q=0
\label{C16}
\ed
if the roots $\alpha$ and $\beta$ are neighboring and $\alpha < \beta$
in a sense of fixed normal ordering of the root system $\Delta_{+}$.

\noindent
(ii) The defining relations (\ref{C2})-(\ref{C5}) may be easily
generalized to the identities
\bn
[e_{n\delta +\alpha_i}, e_{(m+1)\delta -\alpha_i}]=
c\delta_{ij} {e'}_{(n+m+1)\delta}^{(i)}
\label{C17}
\ed
for $n, m \in {\bf Z}$, where $c$ is a constant. The relations
(\ref{C16}) rewritten by means of (\ref{C6}) in terms of the generators
$e_{l\delta \pm \alpha_{i}}^{(i)}$ give us (\ref{D6}).

\noindent
(iii) The formulas (\ref{D5}) define quadratic relations between
the root vectors $e_{n\delta+\alpha_i}$ or between the generators
$e_{n\delta -\alpha_i}$, ($i=1, 2, \ldots, r$).
We write down them explicitly for $U_q(\hat{sl}_2)$ in Appendix A
(see (A.10)-(A.21)).

\setcounter{equation}{0}

\section{The second Drinfeld's realization as twisting of Hopf
structure in $U_q(\hat{g})$}

The second Drinfeld's  realization of the quantum affine algebra
$U_q(\hat{g})$ was originally obtained as a quantization of
a Lie bialgebra structure in the affine Lie algebra $\hat{g}$,
which is equivalent to a presentation of
$\hat{g}\oplus \kappa$ (where $\kappa$ is the Cartan
subalgebra of $g = n_+ \oplus \kappa \oplus n_-$) as a classical double
of the current algebra $\hat{n}_+: =
n_+[t,t^{-1}]\oplus \kappa [t]\oplus {\bf C}d$.
It turned out that the multiplicative structure of this quantization
is isomorphic to $U_q(\hat{g})$ but no any evident connection of
their comultiplication structures was found. Here we make clear
this connection.

The algebra $\hat{n}_+$ may be considered as an image of the Borel
subalgebra $\hat{b}_+$ under a action of a limiting longest element
$w_0$ of the Weyl group $W(\hat{g})$ of $\hat{g}$.
This limiting element $w_0$ does not act on vectors in the Cartan
subalgebra or on the root vectors in $\hat{g}$, nevertheless the
twisting of the Hopf structure in $U_q(\hat{g})$ by $w_0$ is well
defined as a limit of twistings by finite elements of the Weyl group
$W(\hat{g})$ in the FS topology of $T_q(\hat{g}\otimes \hat{g})$
(just as in $T_q^+(\hat{g}\otimes \hat{g})$, see the Section 3).
We prove that Drinfeld's comultiplication
in his second  realization can be obtained as the twisting by $w_0$
of the standard comultiplication in $U_q(\hat{g})$. Analogously to
the Theorem 5.1 we state also that this twisting can be presented as
conjugation by an infinite product of factors of the universal R-matrix.
\begin{theorem}
{\rm (i)}  The expression
\bn
\Delta^{(D)}(a)=(\prod_{\gamma <\delta}R_{\gamma}^{21})^{-1}
\Delta_{q^{-1}}(a)(\prod_{\gamma <\delta}R_{\gamma}^{21})
\label{DT1}
\ed
is well defined element in $T_q(\hat{g}\otimes \hat{g})$
(just as in $T_q^+(\hat{g}\otimes \hat{g}))$ for any fixed element
$a\in U_q(\hat{g})$.

\noindent
{\rm (ii)} The map $\Delta^{(D)}$: $U_q(\hat{g})\rightarrow
T_q(\hat{g}\otimes \hat{g})$ (just as $\Delta^{(D)}$:
$U_q(\hat{g})\rightarrow T_q^+(\hat{g}\otimes \hat{g}))$ is
homomorphism and it satisfies all the axioms of the Hopf algebra.

\noindent
{\rm (iii)} The explicit formulas for $\Delta^{(D)}$ look as follows:
\bn
\Delta^{(D)}(e_{\alpha_i})=e_{\alpha_i}\otimes 1+
k_{\alpha_i}\otimes e_{\alpha_i}-(q-q^{-1})\sum_{m\geq 1}
e_{-m\delta}'^{(i)}k_{\alpha_i}\otimes e_{m\delta +\alpha_i} \ ,
\label{DT2}
\ed
\bn
\Delta^{(D)}(e_{-\alpha_i})= e_{-\alpha_i}\otimes k_{\alpha_i}^{-1}+
1\otimes e_{\alpha_i}- (q-q^{-1})\sum_{m\geq 1}e_{-m\delta -\alpha_i}
\otimes e_{m\delta}'^{(i)}k_{\alpha_i}^{-1} \ ,
\label{DT3}
\ed
\bn
\Delta^{(D)}(e_{n\delta}^{(i)})=e_{n\delta}^{(i)}\otimes 1 +
k_\delta^{-n}\otimes e_{n\delta}^{(i)} \ , \;\;\;\;\;\;
\Delta^{(D)}(e_{-n\delta}^{(i)})=
e_{-n\delta}^{(i)}\otimes k_\delta^n +1\otimes e_{-n\delta}^{(i)} \ .
\label{DT4}
\ed
for $i=1,2,\ldots,$; $n \geq 0$.
In terms of Drinfeld generators the comultiplication $\Delta^{(D)}$ has
the form (\ref{D13})-(\ref{D15}).
\end{theorem}
We shall give a complete proof of this theorem for the case of
$U_q(\hat{sl_2})$ in Appendix B. The general case has no essential
changes. The crucial idea is to look to twistings by powers of
the Lusztig automorphisms that correspond to translations in
the affine Weyl group. It is possible to control the main terms of
twisted coproduct and see what is left in topological limit.

\noindent
{\it Remarks}. (i) We can also obtain the explicit expression
of the right part of (\ref{DT1}) by some direct applications of
q-analogs of the H'Adamard identities (see [KT1]).
Note that only linear terms (or one-step
q-commutators) in the H'Adamard identities give nonzero contributions
in the formula (\ref{DT1}) and corresponding terms appear with a
constant $(q-q^{-1})$ proportional to the "Planck constant" $h$.
In this sense the comultiplication $\Delta^{(D)}$ has quasiclassical
nature.

\noindent
(ii) The technique developed here allows to obtain the connection
between two comultiplications only in one direction. We cannot invert
the procedure and obtain $\Delta_{q^{-1}}$ in $U_q(\hat{g})$
from $\Delta^{(D)}$.

Let  $U_q^{(D)}(\hat{g})$ denote the second Drinfeld realization
of the quantum algebra $U_q(\hat{g})$ with the comultiplication
$\Delta^{(D)}$: $U_q(\hat{g})\mapsto T_q(\hat{g} \otimes \hat{g})$.
A natural question arises whether this Hopf algebra is quasitriangular
and what is the formula for the universal R-matrix for
$U_q^{(D)}(\hat{g})$.
Let us at first remind what is going on with the standard Hopf algebra
structure in $U_q(\hat{g})$. This Hopf algebra is quasitriangular
and for given normal ordering satisfying (\ref{C1}) the universal
R-matrix can be presented in a form (see (\ref{R3})-(\ref{R6})):
\bn
R= R_{+}R_{0}R_{-}K \ ,
\label{DT5}
\ed
where
\bn
R_{+} = \prod_{\alpha <\delta} R_{\alpha} \ , \hsp
R_{-} = \prod_{\delta > \alpha}R_{\alpha} \ ,
\label{DT6}
\ed
\bn
R_0= \exp((q-q^{-1})\sum_{n>0}\sum_{i,j}^{mult}
c_{ij}(n)(e^{(i)}_{n\delta} \otimes e^{(j)}_{-n\delta})) \ .
\label{DT7}
\ed
The products in (\ref{DT6}) are taken over all the real roots located
only on the left side and only on the right one of the imaginary roots
in the normal ordering of $\Delta_{+}(\hat{g})$.

\noindent
By using (\ref{DT1}) and (\ref{R1}) we have that
\bn
\tilde{\Delta}^{(D)}(a)=(R^{(D)})^{-1}\Delta^{(D)}(a) R^{(D)} ,
\hsp \forall\,\,a \in U_{q}(\hat{g}) \ ,
\label{DT8}
\ed
where
\bn
R^{(D)} = R_{0}R_{-}KR^{21}_{+}
\label{DT9}
\ed
and  $R^{(D)}$ may be considered as universal R-matrix for
$U_q^{(D)}(\hat{g})$.

\noindent
Unfortunately, any interpretation of the equality (\ref{DT8}) in
concrete representations is not so simple. Indeed, let $V$ be a
finite-dimensional representation of $U_q(g)$, $V_{z_1}$ and $V_{z_2}$
be corresponding two representations of  $U_q(\hat{g})$ shifted by
$z_1$ and $z_2$ (see [FR] for definitions).
Then the expressions $\Delta^{(D)}(a)(v_{z_1}\otimes
v_{z_2})$, $R_{0}R_{-}K(v_{z_1}\otimes v_{z_2})$ (where $v_{z_1}\in
V_{z_1}$, $v_{z_2}\in V_{z_2})$ are regular for $\mid z_1\mid < \mid z_2
\mid$ and singular for $\mid z_1\mid > \mid z_2 \mid$, and vice versa
the expressions $\tilde{\Delta}^{(D)}(a)(v_{z_1}\otimes v_{z_2})$,
$R_{+}^{21}(v_{z_1}\otimes v_{z_2})$ are regular for
$\mid z_1\mid > \mid z_2\mid$ and singular for
$\mid z_1\mid < \mid z_2 \mid$.

\noindent
We can rewrite the equality (\ref{DT8}) in the following form
\bn
(R_{+}^{21}(\tilde{\Delta}^{(D)})(a)(R_{+}^{21})^{-1}(v_{z_1}\otimes
v_{z_2})=
(R_{0}R_{-}K)^{-1}\Delta^{(D)}(a) R_{0}R_{-}K (v_{z_1}\otimes v_{z_2})
\label{DT10}
\ed
\noindent
with the left side being originally defined for
$\mid z_{1}\mid < \mid z_{2} \mid$ and the right side for
$\mid z_{1}\mid > \mid z_{2} \mid$. (The point is that now both sides
of (\ref{DT10}) have no singularities only on diagonal $z_1=z_2$
and the equality (\ref{DT10}) has rigorous sense).

\noindent
Thus we see that there is no any definite sense for the representations
of $T_q(\hat{q}\otimes \hat{g})$ in $V_{z_1}\otimes V_{z_2}$.
On the other hand, the algebra $T_q^+(\hat{q}\otimes \hat{g})$ acts
on $V_{z_1}\otimes V_{z_2}$
for $\mid z_{1}\mid < \mid z_{2} \mid$ and the algebra
$T_q^-(\hat{q}\otimes \hat{g})$ acts on $V_{z_1}\otimes V_{z_2}$
for $\mid z_{1}\mid > \mid z_{2} \mid$. In this context one can consider
the comultiplication $\Delta^{(D)}$  as a map $U_q(\hat{g})\rightarrow
T_{q}^{+}(\hat{g} \otimes \hat{g})$ and the opposite comultiplication
$\tilde{\Delta}^{(D)}$  as a map $U_q(\hat{g})\rightarrow
T^-_q(\hat{g} \otimes \hat{g})$ and the universal R-matrix $R^{(D)}$ as
the operator: $T_q^-(\hat{q}\otimes \hat{g})\rightarrow
T^+_q(\hat{g} \otimes \hat{g})$. In terms of $V_{z_1}\otimes V_{z_2}$
the operator $R^{(D)}$  has entries being  generalized functions
of $z=\frac{z_1}{z_2}$.

For illustration we consider the concrete example $g = sl_2$.
Let $\rho$ be a two-dimensional representation, then modulo scalar
function (see [KT4])  we have the following formulas
\bn
(\rho_{z_1} \otimes \rho_{z_2}) R_{+}= 1\otimes 1+(\sum_{n\geq 0}z^n)
(e_{12}\otimes e_{21}) \ ,
\label{DT11}
\ed
\[
(\rho_{z_1} \otimes \rho_{z_2}) R_0=
e_{11}\otimes e_{11} + e_{22}\otimes e_{22}+
\]
\bn
+(\exp \sum_{n>0}\frac{q^{2n}-1}{n}z^n)(e_{11}\otimes e_{22})+
(\exp \sum_{n>0}\frac{1-q^{-2n}}{n}z^n)(e_{22}\otimes e_{11}) \ ,
\label{DT12}
\ed
\bn
(\rho_{z_1} \otimes \rho_{z_2}) R_{-}= 1\otimes 1+(\sum_{n\geq 0}z^n)
(e_{21}\otimes e_{12}) \ ,
\label{DT13}
\ed
\bn
(\rho_{z_1}\otimes \rho_{z_2})K=
q^{\frac{1}{2}}( e_{11}\otimes e_{11}+e_{22}\otimes e_{22}) +
q^{-\frac{1}{2}}( e_{11}\otimes e_{22}+e_{22}\otimes e_{11}) \ ,
\label{DT14}
\ed
\bn
(\rho_{z_1} \otimes \rho_{z_2}) R_{+}^{21}=
1\otimes 1+(\sum_{n\geq 0}z^{-n}) (e_{21}\otimes e_{12}) \ ,
\label{DT15}
\ed
and also
\[
(\rho_{z_1} \otimes \rho_{z_2}) R^{(D)} =
(q^{\frac{1}{2}}( e_{11}\otimes e_{11}+e_{22}\otimes e_{22})
+\frac{1-z}{q^{\frac{1}{2}}(1-q^2z)}(e_{11}\otimes e_{22}) +
\]
\bn
+\frac{1-q^{-2}z}{q^{\frac{1}{2}}(1-z)}(e_{22}\otimes e_{11}))
\cdot ( 1 \otimes 1 +\delta(z)
(e_{21}\otimes e_{12}))
\label{DT16}
\ed
where $z:=\frac{{z_1}}{z_{2}}$, $\delta(z):=
\sum_{-\infty}^{\infty} z^n$.

\vspace{12 mm}
\noindent
{\Large{\bf Appendices.}}

\noindent
In this section we exhibit the construction of the Cartan-Weyl basis
and give a complete list of commutation relations between the
Cartan-Weyl generators for $U_q(\hat{sl}_2)$. We also demonstrate
here the proof of the Theorem 8.1 for this case.
\vspace{4 mm}

\noindent
{\large \bf  A. The \CW basis of $U_q(\hat{sl}_2)$.}

\noindent
Let $\alpha$ and $\beta:=\delta -\alpha$ are simple roots for the affine
algebra $\hat{sl}_{2}$ then $\delta=\alpha+\beta$ is a minimal
imaginary root.  We fix the following normal ordering in $\Delta_+$:
$$
\alpha,\;\delta +\alpha, \ldots, \infty\delta +\alpha ,\;
\delta,\; 2\delta, \ldots, \infty \delta,\; \infty\delta -\alpha,\ldots,
2\delta -\alpha ,\;\delta -\alpha \ .
\eqno {\rm (A.1)}
$$
The another normal ordering is the inverse to (A.1):
$$
\delta -\alpha , \;2\delta -\alpha , \ldots , \infty\delta -\alpha ,\;
\delta,\; 2\delta, \ldots, \infty \delta,\; \infty\delta +\alpha,\ldots,
\delta +\alpha , \; \alpha \ .
\eqno {\rm (A.2)}
$$
In accordance with our procedure for the construction of the Cartan-Weyl
basis we put ($n = 1, 2, \ldots$)
$$
e_{\delta}=[e_{\alpha},e_{\delta - \alpha}]_{q} \ ,
\eqno {\rm (A.3)}
$$
$$
e_{n\delta +\alpha}=(-1)^{n}\left( [(\alpha ,\alpha )]_{q}\right)^{-n}
(\tilde{\rm ad}\ e_{\delta})^{n}e_{\alpha} \ ,
\eqno {\rm (A.4)}
$$
$$
e_{(n+1)\delta -\alpha}=\left( [(\alpha ,\alpha )]_{q}\right)^{-n}
(\tilde{\rm ad}\ e_{\delta})^{n}e_{\delta - \alpha} \ ,
\eqno {\rm (A.5)}
$$
$$
e'_{(n+1)\delta}=
[e_{n\delta + \alpha},e_{\delta - \alpha}]_{q}
\eqno {\rm (A.6)}
$$
and then we redefine the imaginary roots $e'_{n\delta}$ by means
of the Schur polynomials:
$$
e_{n\delta}=\sum_{p_{1}+2p_{2}+\ldots +np_{n}=n}
{(q^{-1}-q)^{\sum p_{i}-1}(\sum_{i=1}^{n} p_i -1)!
\over p_{1}!\ldots p_{n}!}
e_{\delta}'^{p_1}e_{2\delta}'^{p_2}\ldots e_{n\delta}'^{p_n} \ .
\eqno {\rm (A.7)}
$$
We take also  $e_{-\gamma}$ = $e_{\gamma}^{*}$,
($\gamma \in \Delta_{+}$).

\noindent
The following formulas are  a total list of the relations for
the Cartan-Weyl generators $e_{\pm \gamma}$, ($\gamma \in \Delta_+$):
$$
k_{\gamma}e_{\pm \gamma'}=
q^{\pm (\gamma,\gamma')}e_{\pm \gamma'}k_{\gamma} \ ,
\hsp \gamma, \; \gamma' \; \in \Delta_+ \ ,
\eqno {\rm (A.8)}
$$
$$
[e_{\gamma},e_{-\gamma}]=\frac{k_{\gamma}-k_{\gamma}^{-1}}{q-q^{-1}} \ ,
\hsp \gamma \neq n\delta \ ,
\eqno {\rm (A.9)}
$$
$$
[e_{n\d},e_{m\d}]=
\d_{n,-m} a(m)\frac{k_{\d}^{n}-k_{\d}^{-n}}{q-q^{-1}} \ ,
\hsp n, \; m \neq 0 \ ,
\eqno {\rm (A.10)}
$$
$$
[e_{n\delta +\alpha},e_{m\delta -\alpha}]_q = e'_{(n+m)\delta} \ ,
\hsp n \geq 0, \; m > 0 \ ,
\eqno {\rm (A.11)}
$$
$$
[e_{n\delta +\alpha},e_{-m\delta - \alpha}]=
-e'_{(n-m)\d}k_{m\delta +\alpha}^{-1} \ ,
\hsp   n > m \geq 0 \ ,
\eqno {\rm (A.12)}
$$
$$
[e_{n\delta - \alpha},e_{-m\delta +\alpha}]=
k_{m\delta-\alpha}e'_{(n-m)\delta} \ ,
\hsp  n > m\ > 0 \ ,
\eqno {\rm (A.13)}
$$
$$
[e_{n\delta +\alpha},e_{m\delta}]= a(m) e_{(n+m)\delta +\alpha} \ ,
\hsp  n \geq 0,\; m > 0,
\eqno {\rm (A.14)}
$$
$$
[e_{n\d -\alpha}, e_{m\d}]=-a(m) e_{(n+m)\d - \alpha} \ ,
\hsp n,\;m > 0 \ ,
\eqno {\rm (A.15)}
$$
$$
[e_{n\d +\a},e_{-m\d}]= a(m) e_{(n-m)\d +\a}k_{\d}^m \ ,
\hsp n \geq m > 0 \ ,
\eqno ({\rm A.16)}
$$
$$
[e_{n\d -\a},e_{-m\d}]=-a(m) e_{(n-m)\d -\a}k_\d^{-m} \ ,
\hsp n \geq m > 0 \ ,
\eqno {\rm (A.17)}
$$
$$
[e_{n\d+\a},e_{(n+2m-1)\d+\a}]_q =
(q_{\a}^2-1) \sum_{l=1}^{m-1} q_{\a}^{-l}
e_{(n+l)\d+\a}e_{(n+2m-1-l)\d+\a} \ ,
\eqno {\rm (A.18)}
$$
\[
[e_{n\d+\a},e_{(n+2m)\d+\a}]_q=
(q_{\a}-1)q_{\a}^{m-1}e_{(n+m)\d+\a}^2 +
\]
$$
+ (q_{\a}^2-1) \sum_{l=1}^{m-1} q_{\a}^{-l}
e_{(n+l)\d+\a}e_{(n+2m-l)\d+\a} \ ,
\eqno {\rm (A.19)}
$$
for any $n\geq 0, \; m > 0$
$$
[e_{(n+2m-1)\d-\a},e_{n\d-\a}]_q =
(q_{\a}^2-1) \sum_{l=1}^{m-1} q_{\a}^{-l}
e_{(n+2m-1-l)\d-\a}e_{(n+l)\d-\a} \ ,
\eqno {\rm (A.20)}
$$
\[
[e_{(n+2m)\d-\a},e_{n\d-\a}]_q =
(q_{\a}-1)q_{\a}^{m-1}e_{(n+m)\d-\a}^2+
\]
$$
+(q_{\a}^2-1) \sum_{l=1}^{m-1} q_{\a}^{-l}
e_{(n+l)\d-\a}e_{(n+2m-l)\d-\a} \ ,
\eqno {\rm (A.21)}
$$
for any $n, \; m > 0$. Here in (A.10), (A.14)-(A.17) the coefficient
$a(m)$ is determined by the formula (\ref{D8}) with $\alpha_i =
\alpha_j =\alpha$.

\noindent
In order to obtain the rest of the relations between root vectors
we have to extend the relations (A.11)-(A.21)
to arbitrary values of $n$. This can be done if we use the circular
generators $\hat{e}_{\pm\gamma}$ and $\check{e}_{\pm\gamma}$
(see (\ref{CW12}), (\ref{CW13})).
More precisely, let
$$
\hat{e}_{n\d+\a} = e_{n\d+\a} \ , \hsp
\hat{e}_{-(n+1)\d +\a}=
-k_{(n+1)\d-\alpha}^{-1}e_{-(n+1)\d+\a},\hsp n\geq 0 \ ,
\eqno ({\rm A.22)}
$$
$$
\check{e}_{-n\d-\a}=e_{-n\d-\a} \ , \hsp
\check{e}_{(n+1)\d-\a}=
-e_{(n+1)\d-\a}k_{(n+1)\d-\a},\hsp n\geq 0 \ .
\eqno {\rm (A.23)}
$$
Then the relations (A.11)-(A.21) transform to the same
formulas where $e_{n\d+\a}$ is replaced everywhere by
$\hat{e}_{n\d+\a}$, and $e_{n\d-\a}$ is replaced by
$\check{e}_{(n+1)\d-\a}$ with the only restriction $n\geq 0$.
Now we have after the conjugation by the Cartan involution $(^*)$
the complete list of the relations for the Cartan-Weyl generators.

\noindent
{\it Remark.} We can observe that the relations (A.14)-(A.17)
may be rewritten in the quadratic form if we rewrite the relations
in terms of $e'_{n\d}$, for instance,
$$
[\hat{e}_{n\d+\a}, e'_{m\d}]=
q_{\a}^{1-m}\hat{e}_{(n+m)\d+\a}+(q_{\a}^2-1)
\sum_{l}^{n-1}q_{\a}^{-l} \hat{e}_{(n+l)\d+\a}e'_{(m-l)\d} \ .
\eqno {\rm (A.24)}
$$

The Drinfeld's generators in the case of $U_q(\hat{sl_2})$ have
the form:
$$
k_c:=k_\delta \ , \hsp \chi_{0}:=k_{\alpha} \ , \hsp
\chi_{n}:= e_{n\delta},
\eqno {\rm (A.25)}
$$
$$
\xi^+_{n}=\hat{e}_{n\delta + \alpha} \ , \hsp
\xi^-_{n}=\check{e}_{n\delta - \alpha} \ ,
\eqno {\rm (A.26)}
$$
for any $n \in {\bf Z}$, and
$$
\phi_{0}=k_{\alpha} \ , \hsp
\phi_{-n}=(q^{-1}-q)k_{\alpha}e'_{-n\delta} \ ,
\eqno {\rm (A.27)}
$$
$$
\psi_{0}=k_{\alpha}^{-1} \ , \hsp
\psi_{n}=(q-q^{-1})k_{\alpha}^{-1}e'_{n\delta}
\eqno {\rm (A.28)}
$$
for $n > 0$.
\vspace{4 mm}
\def\b{\beta}

\noindent
{\large \bf B. The connection between two comultiplications for
$U_q(\hat{sl_2})$. The proof of the Theorem 8.1.}

\noindent
Let $s_\a$ and $s_{\d-\a}$ are the elementary reflections of the Weyl
group of $\hat{sl}_2$. The explicit formulas for
the Lusztig automorphisms $\hat{s}_\a$ and $\hat{s}_{\d-\a}$ in
$U_q(\hat{sl}_{2})$ look as follows:
$$
\hat{s}_{\a}(k_{\a}^{\pm 1}) = k_{\a}^{\mp 1} \ ,
\hsp \hat{s}_{\d-\a}(k_{\d-\a}^{\pm})=
k_{\d-\a}^{\mp 1} \ ,
\eqno {\rm (B.1)}
$$
$$
\hat{s}_{\a}(k_{\d-\a}^{\pm 1}) = k_{\d +\a}^{\pm 1} \ ,
\hsp \hat{s}_{\d-\a}(k_{\a}^{\pm 1})= k_{2\d - \a}^{\pm 1} \ ,
\eqno {\rm (B.2)}
$$
$$
\hat{s}_{\a} (e_{\a}) =
-e_{-\a}k_{\a} \ ,\hsp \hat{s}_{\d-\a} (e_{\d-\a})=
-e_{-\d+\a}k_{\d-\a} \ ,
\eqno {\rm (B.3)}
$$
$$
\hat{s}_{\a} (e_{-\a})=-k_{\a}^{-1}e_{\a} \ ,\hsp
\hat{s}_{\d-\a} (e_{-\d+\a})=-k_{\d-\a}^{-1}e_{\d-\a} \ ,
\eqno {\rm (B.4)}
$$
$$
\hsp
\hat{s}_{\a} (\tilde{e}_{n\d \pm \a})=e_{n\d \mp \a} \ ,  \hsp
\hat{s}_{\d-\a} (e_{(n \mp 1)\d \pm \a})=
\tilde{e}_{(n\pm 1)\d \mp\a} \ ,
\eqno {\rm (B.5)}
$$
$$
\hat{s}_\a (\tilde{e}_{n\d})=e_{n\d} \ , \hsp
\hat{s}_{\d-\a}(e_{n\d})=\tilde{e}_{n\d} \ ,
\eqno {\rm (B.6)}
$$

\noindent
for any integers $n \neq 0$. Here in (B.5) and (B.6)
the root vectors $e_{\gamma}$ are constructed in accordance with
the normal ordering (A.1) and the root vectors $\tilde{e}_\gamma$
in accordance with the inverse normal ordering (A.2).

If we put $\hat{t}_{2\d}:= \hat{s}_{\a} \hat{s}_{\d-\a}$ ($t_{2\d}$ is
a translation in the Weyl group) then we have from
(B.1)-(B.4) the relations
$$
\hat{t}_{2\d}(k_{\a}^{\pm 1})=k_{2\d +\a}^{\pm 1} \ , \hsp
\hat{t}_{2\d}(k_{\d -\a}^{\pm 1})=k_{\d +\a}^{\mp 1} \ , \hsp
\hat{t}_{2\d}(k_{\d}^{\pm 1})=k_{\d}^{\pm 1} \ ,
\eqno {\rm (B.7)}
$$
$$
\hat{t}_{2\d}(e_{\a})=e_{2\d+\a} \ , \hsp
\hat{t}_{2\d}(e_{-\a})=e_{-2\d-\a} \ ,
\eqno {\rm (B.8)}
$$
$$
\hat{t}_{2\d}(e_{\d-\a})=-e_{-\d -\a}k_{\d +\a} \ , \hsp
\hat{t}_{2\d}(e_{-\d+\a})=-k_{\d + \a}^{-1}e_{\d +\a} \ ,
\eqno {\rm (B.9)}
$$
$$
\hat{t}_{2\d}(e_{2\d-\a})=-e_{-\a} k_{\a} \ , \hsp
\hat{t}_{2\d}(e_{-2\d+\a})=-k_{\a}^{-1}e_{-\a} \ ,
\eqno {\rm (B.10)}
$$
$$
\hat{t}_{2\d}(e_{(n\mp 1)\d \pm \a})=
e_{(n \pm 1)\d \pm \alpha} \ , \hsp
\hat{t}_{2\d}(e_{m\d})=e_{m\d} \ ,
\eqno {\rm (B.11)}
$$

\noindent
(for any integers $m \neq 0$ and $n \neq 0, \pm 1 $).

Using general arguments of the Section 8 we have \
$$
(\hat{t}_{2\d}^n\otimes \hat{t}_{2\d}^n) \Delta_{q^{-1}}
(\hat{t}_{2\d}^{-n}(a)) = \left( R_{(2n-1)\d+\a}^{21}\right)^{-1}
\cdots \left( R_{\a }^{21}\right)^{-1} \Delta_{q^{-1}} (a)
 R_{\a }^{21}\cdots R_{(2n-1)\d+\a}^{21} \
\eqno {\rm (B.12)}
$$
for any $a\in U_q(\hat{sl_2})$. Now  we want to investigate
the limits of (B.12) when $n\rightarrow\infty$.
For case  $a=e_{\d}$ we have
$$
\Delta_{q^{-1}}(e_{\d}) =e_{\d} \otimes 1 + k_{\d}^{-1} \otimes
e_{\d} + (q_{\a}-q_{\a}^{-1}) e_{\d -\a}k_\a^{-1} \otimes e_{\a}
\eqno {\rm (B.13)}
$$
and
\[
(\hat{t}_{2\d}^n \otimes \hat{t}_{2\d}^n) \Delta_{q^{-1}}
(\hat{t}_{2\d}^{-n}(e_{\d})) =
\]
$$
=e_{\d}\otimes 1 + k_\d^{-1}\otimes
e_{\d} +(q_{\a} -q_{\a}^{-1}) e_{(-2n+1)\d-\a}k_\a^{-1}\otimes
e_{2n\d+\a} \ .
\eqno {\rm (B.14)}
$$
The last summand tends to zero in the FS topology so we have
$$
{\rm lim}_{n \to \infty} (\hat{t}_{2\d}^n \otimes
\hat{t}_{2\d}^n) \Delta_{q^{-1}}(\hat{t}_{2\d}^{-n}(e_{\d}))
= \Delta^{(D)}(e_{\d}) = e_{\d}\otimes 1 +k_\d^{-1}\otimes e_\d
\eqno {\rm (B.15)}
$$
and analogously for other imaginary root vectors.

\noindent
Now let us consider the real root vectors, for example, $a=e_\a$.
We have $\hat{t}_{2\d}^{-n}(e_{\a})=-k_{2\d -\a}^{-1} e_{-2n\d+\a}$
and have to investigate behavior of the element $\Delta_{q^{-1}}
(e_{-2n\d+\a})$ for large $n$. By induction we see that
$\Delta_{q^{-1}}(e_{-2n\d +\a})$  consists of the following monomials
$e_{-2n\d+\a} \otimes k_{2\d -\a} ^{-1}$, and
$$ a\otimes b =e_{-\a}^{l_0}e_{-\d-\a}^{l_1}\cdots e_{-2m\d-\a}^{l_{2m}}
e_{-\d}^{p_{1}}\cdots e_{-2m\d}^{p_{2m}}\otimes e_{-\d+\a}^{l'_{1}}
\cdots e_{-2m\d+\a}^{l'_{2m}} \ ,
$$
with coefficients from Frac($U_q(\kappa \otimes \kappa) $)
(see Section 3), where $m\geq n $.  Further we have
\[
(\hat{t}_{2\d}^n \otimes \hat{t}_{2\d}^n)(a\otimes b )=
e_{-2n\d-\a}^{l_0}\cdots e_{-2(m+n)\d-\a}^{l_{2m}}
e_{-\d}^{p_{1}}\cdots e_{-2m\d}^{p_{2m}}\otimes e_{(2n-1)\d+\a}^{l'_{1}}
\cdots e_{2(n-m)\d+\a}^{l'_{2m}}
\]
 From weight analysis it is clear that the only nonvanishing terms
in the FS topology (or in the topology of $T^+(\hat{g}\otimes\hat{g})$)
for $\Delta^{(D)}(e_\a)$ are
\[
e_\a\otimes 1\hsp {\rm and} \hsp e_{-m\d}\otimes e_{m\d+\a},
\hsp (m \geq 0) \ ,
\]
with coefficients from Frac($U_q(\kappa \otimes \kappa) $).
After inductive calculations of these coefficients we have (\ref{DT2})
and the statement of the theorem.

\end{document}